\documentclass[10pt,letterpaper]{article}

\usepackage[top=0.85in,left=2.75in,footskip=0.75in]{geometry}

\usepackage{changepage}

\usepackage[utf8]{inputenc}

\usepackage{textcomp,marvosym}

\usepackage{fixltx2e}

\usepackage{amsmath,amssymb}

\usepackage{cite}

\usepackage{nameref,hyperref}

\usepackage[right]{lineno}

\usepackage{microtype}
\DisableLigatures[f]{encoding = *, family = * }

\usepackage{rotating}

\raggedright
\usepackage[aboveskip=1pt,labelfont=bf,labelsep=period,justification=raggedright,singlelinecheck=off]{caption}

\usepackage{xspace}
\usepackage{subcaption}
\usepackage[squaren,cdot]{SIunits}

\usepackage{color}

\makeatletter
\renewcommand{\@biblabel}[1]{\quad#1.}
\makeatother

\date{}

\usepackage{lastpage,fancyhdr,graphicx}
\usepackage{epstopdf}
\fancyheadoffset[L]{2.25in}
\fancyfootoffset[L]{2.25in}
\newcommand{\frechet}{Fréchet\xspace}
\newcommand{\dH}{$\delta_H$\xspace}
\newcommand{\dF}{$\delta_F$\xspace}
\newcommand{\Ca}{C\textsubscript{$\alpha$}\xspace}
\newcommand{\akeco}{AK\textsubscript{eco}\xspace}
\newcommand{\otc}{open $\rightarrow$ closed\xspace}
\newcommand{\cto}{closed $\rightarrow$ open\xspace}
\newcommand{\ctor}{closed $\leftrightarrow$ open\xspace}
\newcommand{\tnd}{$3N$-dimensional\xspace}
\newcommand{\tnmp}{$\theta_\text{NMP}$\xspace}
\newcommand{\tlid}{$\theta_\text{LID}$\xspace}

\newcommand{\qc}{$Q_\text{1ake}$\xspace}
\newcommand{\qo}{$Q_\text{4ake}$\xspace}

\newcommand{\zr}{$\zeta$-$\rho$\xspace}
\newcommand{\tsup}[1]{\textsuperscript{#1}}

\newcommand{\BigO}[1]{\ensuremath{\operatorname{O}\left(#1\right)}}
\newcommand{\mv}[1]{\ensuremath{\mathbf{#1}}} 

\newcommand{\solidrule}[1][0.7cm]{\rule[0.5ex]{#1}{.4pt}}
\newcommand{\dashedrule}{\mbox{%
  \solidrule[1mm]\hspace{1mm}\solidrule[1mm]\hspace{1mm}\solidrule[1mm]\hspace{1mm}\solidrule[1mm]}}

\begin{document}
\vspace*{0.35in}

\begin{flushleft}
{\Large
\textbf\newline{Path Similarity Analysis: a Method for Quantifying Macromolecular Pathways}
}
\newline
\\
Sean L.\ Seyler\textsuperscript{1},
Avishek Kumar\textsuperscript{1},
M.\ F.\ Thorpe\textsuperscript{1,2},
Oliver Beckstein\textsuperscript{1,*},
\\
\bigskip
\bf{1} Department of Physics and Center for Biological Physics,
Arizona State University, Tempe, AZ, United States of America
\\
\bf{2} Rudolf Peierls Centre for Theoretical Physics,
University of Oxford, Oxford, OX1 3NP, United Kingdom
\\
\bigskip

%
%





* oliver.beckstein@asu.edu

\end{flushleft}
\section*{Abstract}
Diverse classes of proteins function through large-scale conformational changes
and various sophisticated computational algorithms have been proposed to enhance
sampling of these macromolecular transition paths. Because such paths are curves in a
high-dimensional space, it has been difficult to quantitatively compare multiple
paths, a necessary prerequisite to, for instance, assess the quality of
different algorithms. We introduce a method named \emph{Path
  Similarity Analysis} (PSA) that enables us to quantify the similarity between
two arbitrary paths and extract the atomic-scale determinants responsible for
their differences. PSA utilizes the full information available in \tnd
configuration space trajectories by employing the Hausdorff or \frechet
metrics (adopted from computational geometry) to quantify the degree of
similarity between piecewise-linear curves. It thus completely avoids relying on
projections into low dimensional spaces, as used in traditional approaches.
To elucidate the principles of PSA, we quantified the effect of path roughness
induced by thermal fluctuations using a toy model system. Using, as an example,
the closed-to-open transitions of the enzyme adenylate kinase (AdK) in its
substrate-free form, we compared a range of protein transition path-generating
algorithms. Molecular dynamics-based dynamic importance sampling (DIMS) MD
and targeted MD (TMD) and the purely geometric FRODA (Framework Rigidity
Optimized Dynamics Algorithm) were tested along with seven other methods
publicly available on servers, including several based on the popular elastic
network model (ENM). PSA with clustering revealed that paths produced by a given
method are more similar to each other than to those from another method and, for
instance, that the ENM-based methods produced relatively similar paths. PSA
applied to ensembles of DIMS MD and FRODA trajectories of the conformational
transition of diphtheria toxin, a particularly challenging example, showed that
the geometry-based FRODA occasionally sampled the pathway space of force
field-based DIMS MD. For the AdK transition, the new concept of a Hausdorff-pair
map enabled us to extract the molecular structural determinants responsible for
differences in pathways, namely a set of conserved salt bridges whose
charge-charge interactions are fully modelled in DIMS MD but not in FRODA. PSA
has the potential to enhance our understanding of transition path sampling
methods, validate them, and to provide a new approach to
  analyzing conformational transitions.

\section*{Author Summary}
Many proteins are nanomachines that perform mechanical or chemical work by
changing their three-dimensional shape and cycle between multiple conformational
states. Computer simulations of such conformational transitions provide
mechanistic insights into protein function but such simulations have been
challenging. In particular, it is not clear how to quantitatively compare
current simulation methods or to assess their accuracy. To that end, we present
a general and flexible computational framework for quantifying transition
paths---by measuring mutual geometric similarity---that, compared with existing
approaches, requires minimal a-priori assumptions and can take advantage of full
atomic detail alongside heuristic information derived from intuition.  Using our
\emph{Path Similarity Analysis} (PSA) framework in parallel with several
existing quantitative approaches, we examine transitions generated for a toy
model of a transition and two biological systems, the enzyme adenylate kinase
and diphtheria toxin. Our results show that PSA enables the quantitative
comparison of different path sampling methods and aids the identification of
potentially important atomistic motions by exploiting geometric information in
transition paths. The method has the potential to enhance our
  understanding of transition path sampling methods, validate them, and to
  provide a new approach to analyzing macromolecular conformational
  transitions.

\section*{Introduction}
Protein function is intimately linked with the mechanistic nature of
conformational transitions---a central problem in computational biophysics is to
determine the function of a protein given its 3D structure~\cite{Yon1998-jp,
  Karplus2005-vz, Henzler-Wildman2007-bp}. Proteins such as enzymes, molecular
motors and membrane transporters behave much like nano-molecular machines that
perform mechanical or chemical work by undergoing conformational transitions
between two or more metastable states. Large scale conformational changes
comprise the slowest frequency motions of a macromolecule and can take place on
the millisecond time scale and beyond. Equilibrium molecular dynamics (MD) is
arguably one of the most robust approaches to simulating macromolecular
dynamics, in large part due to the availability of full atomistic detail
\cite{Dror:2012cr}. It has has been the workhorse tool for studying the protein
structure--function connection \cite{Orozco:2014dq}. However, conformational
transitions are rare events: crossing events in the transition region of phase
space take place much faster than the (waiting) time scales of metastable
equilibria, often by several orders of magnitude.  Equilibrium simulations thus
disproportionately sample metastable states instead of transition events---the
so-called sampling problem---greatly limiting their ability to generate
conformational transition paths~\cite{Schwartz2009-xm}.

Enhanced path-sampling methods and other computational approaches have been
developed to mitigate the sampling problem inherent to macromolecular transition
events, permitting the observation of physics on time- and length-scales
inaccessible to equilibrium MD (see~\cite{Lei2007-hq, Yang2008-bi, Chng2008-wc,
  Zuckerman2011-ts, Christen2008-if, Seyler2014-uu} for reviews). Conformational
sampling in trajectory-based (i.e., dynamics-based)
methods~\cite{Schlitter1994-xz, Voter1997-lc, Woolf1998-xh, Sugita1999-de,
  Laio2002-qx, Hamelberg2004-ot, Kubitzki2008-qy, Barnett2009-xl, Abrams2013-hh}
is accelerated by reducing the computational cost per time step and/or by
minimizing the total number of time steps needed for
sampling~\cite{Zuckerman2011-ts}. Non-dynamical approaches can be roughly
divided into the class of minimum (free) energy path (MEP/MFEP)
methods~\cite{Bolhuis2002-hj, E2005-vp, Maragliano2006-dy, Van_der_Vaart2007-zp,
  Pan2008-eg, Jonsson1998-dz, Henkelman2000-fm, Henkelman2000-jy,
  Fischer1992-tp}, including elastic network model (ENM)
approaches~\cite{Franklin2007-mz, Tirion1996-qg, Bahar1997-nb, Atilgan2001-qf,
  Maragakis2005-me}, and prior-information/geometry-based
algorithms~\cite{Cortes2005-hw, Seeliger2007-ka, Raveh2009-wa,
  Farrell2010-wh}. A large number of the aforementioned methods overlap
algorithmically or are similar in spirit; many are also directly amenable (or
can be adapted) to performing free energy calculations.  Presently, however, the
full extent to which such coarse-grained (CG) or biased MD approaches can
replicate physical transition ensembles is unknown, especially given the
diversity of physical assumptions of the various models. Thus, tools aiding more
rigorous inspection of the capabilities and effectiveness of path-sampling
methods are needed. In a more general sense we need a means to compare the
protein motions, i.e.\ the transition paths, in an unbiased manner that makes
use of all the available structural information.

\subsection*{Approaches to transition path analysis}
Conformational transition paths are represented by sequences of (snapshots of)
conformers in \tnd configuration space, making it difficult to examine---both
visually and quantitatively---their character without resorting to
dimensionality reduction in a collective variable (CV) space. Native contacts
analysis (NCA), for example, is a general approach frequently used to
characterize protein folding pathways~\cite{Best2013-na} and
enables dimensionality reduction via a projection onto 2D native contacts
(NC) space. NCA has the property that structural contacts are defined
without reference to another structure, making NC space projections
particularly useful when good reaction coordinates are not known a priori.
Another common approach is principal component analysis (PCA), a tool that
can be used to visualize conformational dynamics in a lower-dimensional
subspace spanned by several principle components (PCs)~\cite{Balsera1996-yo,
Kitao1999-jr}. An important aspect of PCA is that motion along PCs can be
viewed in real space, helping make complicated dynamical motions visually
tractable.

Using NCA, PCA or other CV approaches cannot, however, guarantee that important
dynamical motions will be captured in the projections---whether (and what)
dynamical information is lost depends on the projection itself. It is clear that
a quantitative method that can examine a full \tnd trajectory would help
mitigate biases inherent to selecting a coordinate projection. We propose a
general computational method named \emph{Path Similarity Analysis} (PSA) to
quantitatively compare $3N$-dimensional macromolecular transition paths, which
is based on the idea of measuring the geometric similarity between pairs of
paths using path similarity metrics. Based on distances between paths,
  trajectories are then clustered by similarity. The structural determinants
  responsible for the difference between any two trajectories are extracted at
  the atomic level by exploiting properties of the underlying metric. Here we introduce
the PSA approach, examine its suitability, performance, and limitations as a
computational approach to quantifying path similarity and
apply it to a toy system and
conformational transitions of two proteins.

\subsection*{Path metrics for measuring transition path similarity}
Path similarity analysis (PSA) exploits the properties of a (path) metric
function, $\delta$, that measures a distance between a pair of piecewise-linear
or polygonal curves, i.e., an ordered set of vertices connected by edges. A
metric $\delta$ applied to curves $A$, $B$, $C$ has the properties
\begin{subequations}
\begin{align}
    \delta(A,B) &\geq 0 \label{eq:first}\\
    \delta(A,B) &= 0 \iff A=B \label{eq:second}\\
    \delta(A,B) &= \delta(B,A) \label{eq:third}\\
    \delta(A,C) &\leq \delta(A,B) + \delta(B,C). \label{eq:fourth}
\end{align}
\end{subequations}
In particular, Eq.~\ref{eq:second}, the identity property, is essential since
it implies that, given two curves $A$ and $B$, if $B$ were to be continuously
deformed so as to monotonically decrease the distance $\delta(A,B)$, then
$\delta(A,B)\rightarrow 0$ as $B\rightarrow A$. That is, two curves must become
identical as their mutual distance approaches zero so that decreasing values of
$\delta$ correspond to increasing similarity. The other properties
---non-negativity (Eq.~\ref{eq:first}), commutativity (Eq.~\ref{eq:third}) and
triangle inequality (Eq.~\ref{eq:fourth})---guarantee that $\delta$ behaves in
the same way as any other metric usually used in structural comparisons (such as
root mean squared distance) even though it compares whole paths and not just
individual conformations.

PSA does not require the use of true metrics and can be
used with any path distance function or other dissimilarity measure where
only Eqs.~\ref{eq:first}--\ref{eq:third} are satisfied. The triangle
inequality (Eq.~\ref{eq:fourth}), which is a generalization of the transitive property, says that
when two objects, $A$ and $B$, in some metric space, are each close to a third
object, $C$, in the same space, then $A$ is close to $B$ in the sense that
the triangle inequality, $d(A,B) \leq d(A,C) + d(B,C)$, provides an upper
bound on their distance apart. The triangle inequality is therefore important
when comparing more than two objects, which is the common scenario when
analyzing many conformational transitions. Although in the following we only
consider true metrics, we also explore several distance functions that
violate the triangle inequality in \nameref{S1_Text}. In the main part of
this study, we consider two candidates for $\delta$---the Hausdorff
metric~\cite{Huttenlocher1993-rr, Alt1995-dh, Alt2008-lg} and the discrete
\frechet metric~\cite{Frechet1906-ih, Alt1995-mc}---and illuminate situations
where one might be selected in favor of the other. Given two paths as input,
both metrics locate two points, one per path, corresponding to some notion of a
maximal deviation between the paths. An important property of these metrics is
that they are sensitive only to path geometry; they are insensitive to dynamical
motions and associated physical time scales along paths. We provide a brief
overview of these two path metrics in the context of conformational transitions.

\paragraph{Hausdorff metric.}
We start with a \tnd configuration space containing two paths $P$ and $Q$
represented, respectively, as sequences of conformations $\{(p_k)_{k=1}^n
\mid p_k\in\mathbb{R}^{3N}, k=1,\dots,n\}$ and $\{(q_k)_{k=1}^m \mid q_k\in
\mathbb{R}^{3N}, k=1,\dots,m\}$. The \emph{Hausdorff distance} is defined as
\begin{equation}\label{eq:dh}
  \delta_{H}(P,Q) = \max\left\{\delta_h(P\mid Q),\delta_h(Q\mid P)\right\},
\end{equation}
where
\begin{equation}\label{eq:direct-dh}
  \delta_{h}(P\mid Q) = \max_{p\in P}\min_{q\in Q} d(p,q)
\end{equation}
is the \emph{directed Hausdorff distance} from $P$ to $Q$, and $d$ is a distance
metric on $\mathbb{R}^{3N}$ (measuring point
distances)~\cite{Huttenlocher1993-rr}; the vertical bar ($P\mid Q$) emphasizes
that $\delta_{h}(P\mid Q)$ is not commutative. The function $\delta_{h}(P\mid
Q)$ selects the point $p*\in P$, among all points in $P$, with the most distant
nearest neighbor $q*\in Q$ (as measured by $d(p*,q*)$). In the language of
conformational transitions, we interpret $d(p,q)$ as a putative structural
similarity measure between conformers $p$ and $q$, so that for some conformer
$p_k\in P$, its structural ``nearest neighbor'' in $Q$ is given by $\min_{q\in
Q} d(p_k,q)$. Thus, $\delta_{h}(P\mid Q)$ is the distance $d$ associated with
the conformer in $P$ having the \emph{most distant} or \emph{least similar}
nearest neighbor (in $Q$). The Hausdorff distance between $P$ and $Q$,
$\delta_H(P,Q)$, is therefore the distance associated with the point---\emph{of
all points in $P$ and $Q$}---with the least similar nearest neighbor, and
implies that all points have a nearest neighbor that is at most $\delta_H(P,Q)$
away.

\paragraph{\frechet metric.}
Unlike the Hausdorff metric, \frechet metrics are sensitive to the orientation
(i.e., directionality) of paths; real transition paths are inherently
directional which in principle makes \frechet metrics superior to the Hausdorff
metric. Informally, the \emph{continuous \frechet distance} can be visualized
by considering a man walking on a path $P$ and his dog on another path
$Q$~\cite{Alt1995-mc}. Both start at the initial points of their respective
paths, and they are imagined to be connected by an elastic leash that remains
taught so as to measure the distance separating them at all times. We then allow
the man and dog to move independently on their respective paths under the
condition that each progresses in a monotonic fashion (i.e., no backward steps)
from start to finish. The \frechet distance between $P$ and $Q$ is
then defined as the length of the shortest leash necessary
for the man and dog to move along their respective paths from beginning to end
according to the aforementioned constraints.  Formally, for two continuous
curves $P: [a_{0},a_{1}] \rightarrow \mathbb{R}^{3N},\ a_{0} < a_{1}$ and $Q:
[b_{0},b_{1}] \rightarrow \mathbb{R}^{3N},\ b_{0} < b_{1}$ that are
parameterized with a real parameter, the continuous \frechet distance
corresponds to finding two specific continuous and monotonous parameterizations
$\alpha: [0,1]\rightarrow[a_{0}, a_{1}]$ and $\beta:
[0,1]\rightarrow[b_{0},b_{1}]$ (the ``schedules'' of the man and the dog along
their paths) so that the largest point distance $d$ for a given set of
parameterizations is minimized~\cite{Alt1995-mc},
\begin{equation}
  \label{eq:contFrechet}
  \delta_{F}(P,Q) = \min_{\alpha, \beta} %
      \max_{t\in[0,1]}d\Big(P\big(\alpha(t)\big), Q\big(\beta(t)\big)\Big).
\end{equation}
Algorithms exist to solve this difficult problem in \BigO{nm\log nm} time for
polygonal curves (where $n$ and $m$ are the number of vertices in each
curve)~\cite{Alt1995-mc} and various faster approximate solutions have been
suggested~\cite{Driemel2012-ji, Har-Peled2014-nx}.

In this paper, however, we exclusively use the \emph{discrete \frechet
  distance}, $\delta_{dF}$, with the algorithm outlined by~\cite{Eiter1994-wz}
as it is simpler and faster to compute (in \BigO{nm} time) than its continuous
counterpart, \dF. The formal definition of $\delta_{dF}$ considers two polygonal
curves $P$ and $Q$ that are defined respectively by $n$ and $m$ ordered points
in a metric space $(V,d)$ for some metric $d$. Let the corresponding sequence of
endpoints of the line segments of $P$ and $Q$ be respectively defined as
$\sigma(P)=(p_1,\dots,p_n)$ and $\sigma(Q)=(q_1,\dots,q_m)$. In the product
space $\sigma(Q,P) \equiv \sigma(P)\times\sigma(Q)$, we define a \emph{coupling}
between two polygonal curves $P$ and $Q$ as a sequence,
\begin{equation}
  C(P,Q) \equiv (p_{a_1},q_{b_1}),(p_{a_2},q_{b_2}),\dots,(p_{a_L},q_{b_L}),
\end{equation}
of $L$ unique pairs of points (i.e., number of links) satisfying the following
conditions: (1) The first/last pairs correspond to the first/last points of the
respective paths ($a_1=b_1=1$, $a_L=n$ and $b_L=m$); (2) at least one point on
a path (for a pair of points, one per path) must be advanced to its successive
point, i.e.,\ ($a_{i+1}=a_i$ and $b_{i+1}=b_i+1$) or
($a_{i+1}=a_i+1$ and $b_{i+1}=b_i$) or ($a_{i+1}=a_i+1$ and $b_{i+1}=b_i+1$)
for all $i=1,\dots,L$. The largest distance between a pair of points
$(p_{a_i},q_{b_i})$ for a given coupling $C$ defines the coupling distance
\begin{equation}
  \label{eq:coupl_dist}
  \|C\| \equiv \max_{i=1,\ldots,L} d(p_{a_i},q_{b_i}).
\end{equation}
Given the space of all possible couplings between $P$ and $Q$, $\Gamma_{P,Q}$,
the \emph{discrete \frechet distance} between $P$ and $Q$ is the minimum
coupling distance among all couplings in $\Gamma_{P,Q}$:
\begin{equation}
  \label{eq:dF}
  \delta_{dF}(P,Q) = \min_{C\in\Gamma_{P,Q}}{\|C\|}.
\end{equation}

\begin{figure}[t]
  \centering
  \includegraphics[]{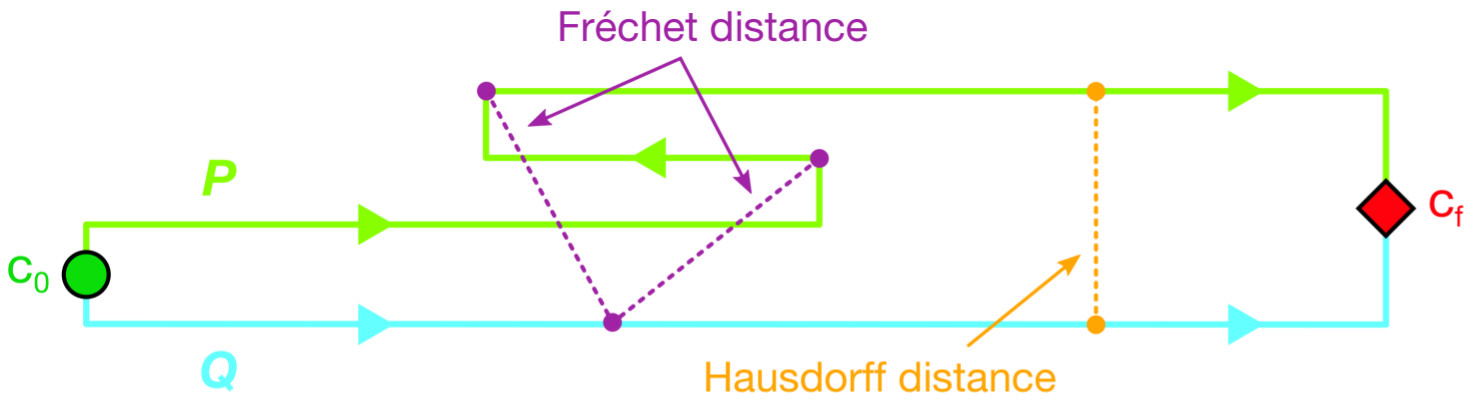}
  \caption{Two paths $P$ (green) and $Q$ (cyan) begin at state $c_0$ and end at
    state $c_f$ with directionality indicated by the arrows. The \frechet
    distance $\delta_F$ and Hausdorff distance $\delta_H$ are given by the
    lengths of the purple and orange lines, respectively. The purple lines are the
    same length and correspond to the minimally stretched \frechet ``leash'';
    the orange line spans a pair of points separated by the Hausdorff distance (only
    one is shown because in this case there are infinitely many pairs of points
    with the same $\delta_{H}$). Due to the backtracking of path $P$ toward state
    $A$, combined with the monotonicity (no-backward-movement) constraint of the
    \frechet metric, $\delta_{F} > \delta_{H}$.}
  \label{fig:frechethaus}
\end{figure}

The continuous \frechet distance constitutes a lower bound on the discrete
\frechet distance, $\delta_{F} \leq \delta_{dF}$, because $\delta_{F}$ accounts
for points along the (straight) edges connecting the vertices, whereas
$\delta_{dF}$ only takes the vertices themselves into consideration~\cite{Eiter1994-wz}.
Furthermore, if we define the maximum edge length for a polygonal curve $P$ to
be the largest distance between consecutive points in $P$, $d_\text{max}(P) \equiv
\max_{i=1,\ldots,p-1}d\left(p_i,p_{i+1}\right)$, we can set an upper bound on
$\delta_{dF}$ given two polygonal curves $P$ and $Q$ so that $\delta_F\leq
\delta_{dF}(P,Q) \leq \delta_F(P,Q) + \max \{d_\text{max}(P),
d_\text{max}(Q)\}$~\cite{Eiter1994-wz}. Thus, $\delta_{dF}$ differs from
$\delta_{F}$ by no more than the longest edge among both paths and, to good
approximation, $\delta_{dF}\approx\delta_{F}$ for typical trajectories with
regularly spaced conformations. Hereafter we refer to the discrete
\frechet distance as simply the \frechet metric (distance) with symbol \dF{}
for brevity. The \frechet distance is bounded from below by the Hausdorff
distance for any given pair of piecewise-linear curves~\cite{Alt2001-wh}
($\delta_{F} \geq \delta_{H}$) because for convex polygonal curves the \frechet
and Hausdorff distances are equal~\cite{Buchin2008-tv} while for other path
geometries the \frechet distance can become arbitrarily larger than the Hausdorff
distance~\cite{Driemel2012-ji}. In the case of macromolecular trajectories, the
case of backtracking appears particularly relevant because of its conceptual
link to a random walk and its connection to thermal fluctuations. If one path
runs backward along some portion relative to another path, the \frechet
distance will increase with the extent of the backtracking, whereas the
Hausdorff distance will be unaffected since it ignores the direction of path
traversal (Fig.~\ref{fig:frechethaus}).

\paragraph{Measuring structural similarity.}
Both the Hausdorff and \frechet distances defined in Eq.~\ref{eq:dh} and
Eq.~\ref{eq:dF}, respectively, are defined in terms of a point metric $d(p,q)$
on \tnd configuration space that measures the distance (i.e., similarity)
between conformations $p$ and $q$. We employ the root mean square distance
(rmsd) defined in the usual way as
\begin{equation}\label{eq:rmsd}
  d_\text{RMS}(p,q) = \sqrt{\frac{1}{N}
    \sum_{i=1}^{3N}\left(p_{i}-q_{i}\right)^2},
\end{equation}
where $N$ is the number of atoms, and $\{p_i\}_{i=1}^{3N}$ and
$\{q_i\}_{i=1}^{3N}$ define the configuration space coordinates of
conformations $p$ and $q$, respectively.

It should be noted that Hausdorff and \frechet metrics can be defined
in terms of other point metrics to measure and thus emphasize
different aspects of macromolecular structure or topology. For example,
one could choose to measure the similarity of two protein conformers by
quantifying the percentage of shared contacts. Another promising
approach may be to integrate information-based metrics used for measuring
the similarity of protein ensembles~\cite{Lindorff-Larsen2009-wm}.
In this paper, we exclusively used the best-fit rmsd as the point metric
due to its simplicity and widespread use, helping to connect with familiar
intuitions and avoid obfuscating the examination of the path metrics themselves.

\paragraph{Previous studies and alternative approaches.}
The Hausdorff metric has found applications in image
comparison~\cite{Huttenlocher1993-rr}, while the orientation-dependent \frechet
metric has been used for handwriting recognition and searching handwritten
documents~\cite{Sriraghavendra2007-lv}, and comparing trajectories of moving
objects in geographic information systems~\cite{De_Berg2011-kw}. Both metrics
have also found applications in biology for protein structure
alignment~\cite{Zhu2007-fl, Jiang2008-fv} and protein homology
analysis~\cite{Panchenko2004-gz, Panchenko2005-qa}.

To our knowledge, the Hausdorff and \frechet metrics have not been widely used
as general tools to quantify macromolecular pathways. However,
  recently two studies employed \frechet distances to assess convergence of
  transition paths to an optimal path. Jiang et al.~\cite{Jiang2014-zl} used the
  same discrete \frechet metric as used in this study to assess the convergence
  of a swarms-of-trajectories string method.  Dickson et
  al.~\cite{Dickson2012-dm} employed a variation of the discrete \frechet
  distance where the coupling distance was defined as the average distance
  between all pairs in a coupling (instead of the maximum distance as in
  Eq.~\ref{eq:coupl_dist}); this \emph{discrete average \frechet} distance was
  used in combination with an adaptive biasing force method to assess the
  convergence to an optimal path in an a priori CV space and was found to
  produce easier-to-read results by reducing statistical noise compared to the
  conventional metric. We explore this metric in more detail in \nameref{S1_Text}
  along with a type of average Hausdorff distance. Protein
folding pathways have been compared quantitatively but not with Hausdorff or
\frechet metrics. Several such studies utilized native contacts-based path
(dis)similarity measures~\cite{Gin2009-ho, Lenz2009-za, Graham2011-xp,
  Lindorff-Larsen2011-wr}. In particular, both Graham et
al.~\cite{Graham2011-xp} and Lindorff-Larsen et
al.~\cite{Lindorff-Larsen2011-wr} used dissimilarity scores to assign individual
paths to folding pathways using clustering. Different methods to sample
  conformational transitions were compared by Huang et al.~\cite{Huang2009-ws},
  who compared the original targeted MD (TMD) algorithm \cite{Schlitter1994-xz}
  with a harmonic restraint variation of TMD (also known as ``steered MD'' (SMD)
  or ``restrained TMD'' (rTMD) \cite{Ferrara2000-py})---and biased MD (BMD)
  approaches, and Ovchinnikov and Karplus~\cite{Ovchinnikov2012-pf}, who
  analyzed the free energy profiles along the transition tubes surrounding the
  paths produced by several TMD variants.

The use of the \frechet and Hausdorff path metrics on transition paths
  itself is not new; however, their application as general-purpose tools for
  quantitatively analyzing and comparing ensembles of transition paths---and
  extracting the molecular-scale determinants that dictate their
  differences---is, to our knowledge, novel. A particularly important advantage
  of the Hausdorff and \frechet metrics is that they do not require a choice of
  progress variable, unlike metrics based on binning trajectory snapshots to
  compute path rmsds. While we emphasize that PSA suggests a general approach to
  quantitative transition path analysis using different structural and path
  metrics, we restricted our study to the Hausdorff and \frechet path metrics
  implemented with the rmsd (as a structural similarity metric) to demonstrate
  the viability of a basic approach. We attempted to keep the underlying principles of PSA in view
to engender future PSA-based analyses (such as quantifying putative reaction
coordinates) and stress that this study does not purport to exhaust all
applications of PSA, nor represent an optimized application. Other path
metrics---e.g., \frechet with speed limits, direction-based
\frechet~\cite{Maheshwari2011-go}, or \frechet with shortcuts for the analysis
of noisy data~\cite{Driemel2013-zm}---may offer advantages in carrying out
various analyses. The Hausdorff distance can be generalized as well to measure,
for instance, distances between surfaces (instead of 1D
curves)~\cite{Barton2010-dj}. The multitudinous permutations that can be
selected among the various path metrics, structural similarity metrics,
clustering algorithms, etc.\ make PSA a flexible tool for trajectory
  analysis.

\subsection*{Model systems}
\begin{figure}[bt]\label{fig:transitions}
\begin{adjustwidth}{-2.0in}{0.0in}
  \centering
  \includegraphics[]{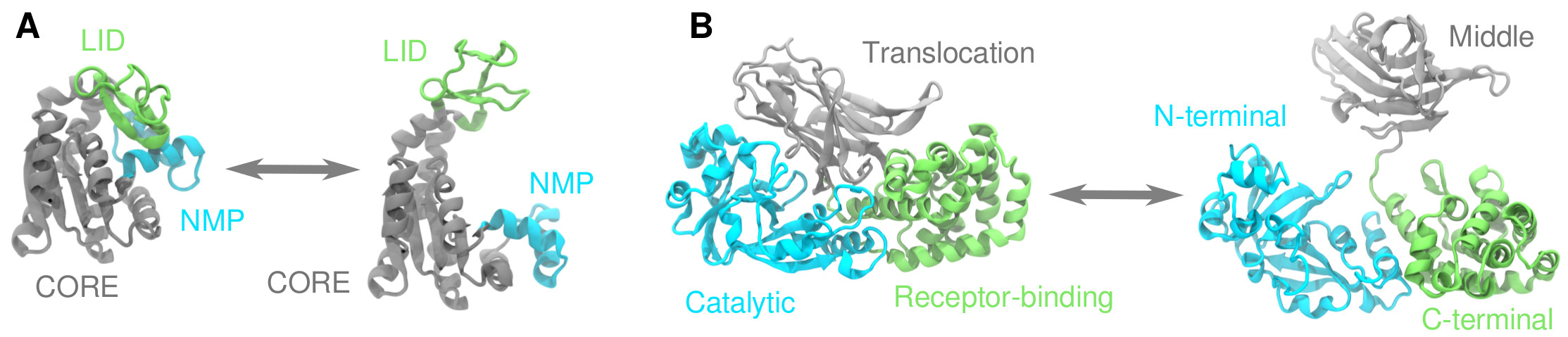}
  \caption{(A) The \ctor transition for adenylate kinase (AdK) involves the
    hinge-like motion of the LID (green) and NMP (cyan) domains about the
    relatively stable CORE (gray). (B) Diphtheria toxin (DT) can exist in two
    different crystallographic conformations that are connected by a \ctor
    transition involving the unravelling and swinging-open of the Translocation
    (T or Middle) domain (gray), about the Catalytic (C or N-terminal; green)
    and the Receptor-binding (R or C-terminal; cyan) domains.}
  \label{fig:combotrans}
\end{adjustwidth}
\end{figure}

To investigate the applicability of the Hausdorff and \frechet metrics to the
problem of quantifying transition paths, we generated trajectories using an
abstract toy system and we simulated conformational transitions of two globular
proteins, the enzyme adenylate kinase (AdK) in its ligand-free form and
diphtheria toxin (DT). The toy model was designed to gain an intuition for the
path metrics and their applicability to highly fluctuating paths in high
dimensions. AdK's closed/open transition (Fig.~\ref{fig:combotrans}A) is a
standard test case that captures essential features of general conformational
changes in proteins~\cite{Seyler2014-uu}. Alongside AdK in our analysis of
transition ensembles, we also examined \cto DT transitions
(Fig.~\ref{fig:combotrans}B), which serves as a more challenging example due to
the difficultly of capturing the putative unfolding and refolding required for
conformational change~\cite{Krebs2000-di}.

AdK is divided into three domains: the ATP-binding (or ``LID'') domain, residues
122-159 in the mesophilic \emph{Escherichia coli} sequence (\akeco), and the
AMP-binding (or ``NMP'' or ``AMPbd'') domain, residues 30-59, move relative to
the CORE domain~\cite{Schulz1990-hp, Gerstein1993-jg, Vonrhein1995-yq,
  Sinev1996-xp, Muller1996-gi} around conserved
hinges~\cite{Henzler-Wildman2007-fb} (Fig.~\ref{fig:combotrans}A). The
conformational change can occur in the ligand-free (apo) state as demonstrated
in multiple experimental studies~\cite{Henzler-Wildman2007-fb, Shapiro2006-xn,
  Hanson2007-wf, Aden2007-kk} and corroborated by computational analyses
(reviewed by Seyler and Beckstein~\cite{Seyler2014-uu}).  Therefore, the apo
\akeco enzyme is a particularly suitable model system for studying general
conformational transitions~\cite{Seyler2014-uu}. We produced transition paths
between an open conformation of AdK [represented by chain A of PDB id
4AKE~\cite{Muller1996-gi} from the Protein Data Bank~\cite{Berman2000-eo}
(PDB)], and a closed conformation (chain A of 1AKE~\cite{Muller1992-bq} with
ligand removed).

DT is believed to undergo a transition from an inactive closed conformation to
an active open one, which includes a 180$^{\circ}$ rotation of a mobile
domain~\cite{Bennett1994-hn} (Fig.~\ref{fig:combotrans}B). An open conformation
was captured in a domain-swapped dimeric structure~\cite{Bennett1994-um} and
compared to the closed monomeric structure~\cite{Bennett1994-im}. DT is divided
into three domains, with the translocation (T) domain, residues 179-379, being
responsible for the majority of the opening and unrolling conformational motion
about the receptor-binding (R) domain, residues 380-535, and the catalytic (C)
domain, residues 1-178. The conformational transition of a DT monomer was
simulated previously and considered challenging for simulation
methods~\cite{Krebs2000-di, Farrell2010-wh}. We simulated transition pathways of
DT between a closed and open conformation based on chain A from the monomeric
structure (PDB id: 1MDT~\cite{Bennett1994-im}) and chain A from the
domain-swapped dimeric structure (PDB id: 1DDT~\cite{Bennett1994-um}),
respectively.

\section*{Methods}
In the following we define the PSA approach as implemented in this study (using
the metrics described in the Introduction), and we also summarize several
alternative approaches to analyzing transitions that we employed alongside PSA
for comparison. We describe how a range of conformational transition paths was
generated to supply a variety of contexts in which to test PSA.

Molecular images were created with VMD \cite{Humphrey1996aa} and the Bendix
plugin \cite{Dahl:2012ap}. Graphs were plotted with the Python libraries
matplotlib \cite{Hunter:2007aa} and seaborn \cite{Waskom2014:aa}, in particular
its implementation of violin plots \cite{Hintze:1998tw}.

\subsection*{Characterizing transition paths}
\paragraph{Path similarity analysis (PSA).}
The Hausdorff metric, $\delta_{H}$, and the discrete \frechet metric,
$\delta_{dF}$, defined in Eq.~\ref{eq:dh} and Eq.~\ref{eq:dF}, respectively,
were computed as described in the Introduction. Further details on the numerical
implementation are provided in \nameref{S2_Text}. Both metrics are implemented
as part of the MDAnalysis Python package~\cite{Michaud-Agrawal2011-yg} in the
module \texttt{MDAnalysis.analysis.psa}, which is available as open source at
\url{www.mdanalysis.org} under the GNU General Public License 2.

To analyze a set of $N$ paths, we compute the $N(N-1)/2$ unique
pairwise Hausdorff and \frechet distances. To present the data
efficiently, we levied the versatility of hierarchical
clustering~\cite{Xu2008-mg} along with the visual power of a heat
map-dendrogram representation to present a quantitative approach to
visualizing the similarities of collections of paths. In agglomerative
hierarchical clustering, objects are linked with similar objects to
form growing clusters in a bottom-up approach. The similarity between
two objects is defined by a metric, while the similarity of clusters
(i.e., sets of objects) is uniquely determined by a linkage criterion
that computes cluster similarity as a function of the pairwise
similarities of the objects comprising each cluster.

Using the Hausdorff and \frechet metrics as similarity measures, we employed
Ward's method~\cite{Ward1963-mp} in conjunction with agglomerative hierarchical
clustering as implemented in the SciPy Python package \cite{SciPy}. The Ward
linkage criterion specifies a minimum variance criterion that minimizes the total
intra-cluster variance. In light of the focus of this paper, we restrict our
study to hierarchical clustering using primarily Ward linkage---details
regarding this restriction are provided in \nameref{S3_Text} in the Supporting
Information along with other relevant considerations in using cluster analysis
to facilitate PSA.

\paragraph{Native contacts analysis (NCA).}
For consistency with other methods used in this paper, we define a contact to be
a residue pair whose \Ca atoms are separated by a distance smaller than
$\unit{8}{\angstrom}$. A \emph{native contact} is a contact present in a
reference structure. Given a transition path, the fraction of native contacts
$Q$~\cite{Shakhnovich1991-ae} is the fraction of contacts in a native structure
that are present in a transition structure. We then define, for any intermediate
conformer in a transition, $Q_1$ and $Q_2$ as the fractions of native contacts
with respect to an initial and final structure, respectively. Transition
paths are projected onto 2D $Q_1$-$Q_2$ (NC) space by parametrically plotting
the percentage of contacts relative to the initial and final states.

\begin{figure}[]
\begin{adjustwidth}{0.5in}{0.5in}
  \centering
  \includegraphics[]{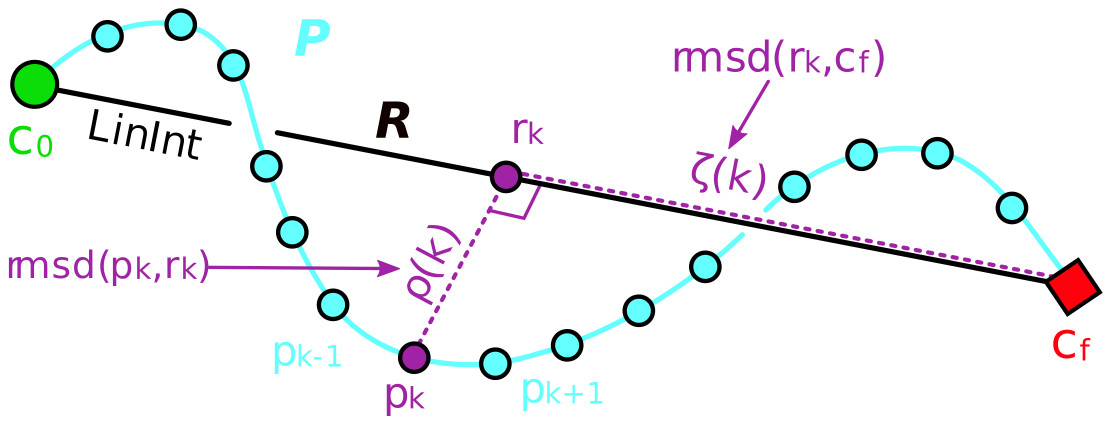}
  \caption{A hypothetical transition pathway $P$ (cyan line) in a
    3D configuration space composed of a discrete number of conformer snapshots
    (cyan circles) connects an initial state (green circle), $c_0$, and final
    state (red diamond), $c_f$. The reference path $R$ (black line) is represented
    by LinInt. Each snapshot $p_k$ is associated with its projection, $r_k$, on
    $R$; the progress, $\zeta(k)$, is the rmsd between $r_k$ and $c_f$ (dashed
    purple line along $R$) and the displacement, $\rho(k)$, is the rmsd between $p_k$ and
    $r_k$ (dashed purple line perpendicular to $R$).}
  \label{fig:pla}
\end{adjustwidth}
\end{figure}

\paragraph{Comparison with a linearly interpolated path.}
A simple way to quantify the geometry of a single transition path is to measure
its orthogonal separation, $\rho$, from a reference path as a function of
progress, $\zeta$, along the reference path (Fig.~\ref{fig:pla}). In this way,
any transition path can be projected in a 2D space depicting ``displacement''
($\rho$) versus ``progress'' ($\zeta$) relative to a reference path. We selected
naive linear interpolation (LinInt) to serve as a zeroth-order reference
transition path. Note that, in comparison with PSA, this approach necessitates
defining an explicit progress measure in the form of a reference path---which
may not be appropriate beyond relatively simple examples like the AdK
transition---and is furthermore not amenable to direct pairwise comparisons
among a large ensemble of transition paths.

Given two boundary conformations $\{c_0,c_f\}\in\mathbb{R}^{3N}$ in \tnd
configuration space with reference path $R$ embedded in $\mathbb{R}^{3N}$ (that
linearly interpolates $c_0$ and $c_f$), and a piecewise-linear (transition) path
$P$ embedded in $\mathbb{R}^{3N}$ and composed of a sequence of conformations,
$(p_k)_{k=1}^m$, where $m$ is the number of time steps, we compute for each
$p_k$: (1) the rmsd between $p_k$ and its orthogonal projection onto $R$, $r_k$,
\begin{equation}
  \rho(k) = d_\text{RMS}(p_k,r_k),
\end{equation}
and (2) the rmsd between $r_k$ and final state $c_f$,
\begin{equation}
  \zeta(k) = d_\text{RMS}(r_k,c_f)
\end{equation}
(see Fig.~\ref{fig:pla}). A transition path can then be projected onto \zr
space by parametrically plotting $\zeta(k)$ versus $\rho(k)$ for all
values of $k$. For a path beginning at $r_0=c_0$, the rmsd to the
final structure is given by the rmsd between the initial and final
states, $\zeta(0) = d_\text{RMS}(c_0,c_f)$, while the rmsd for a path
ending at $r_m=c_f$ is $\zeta(m) = d_\text{RMS}(c_f,c_f) = 0$.

Defining $\rho$ using the rmsd permits a close connection with PSA in the
following way: the maximal rmsd of a path $P$ from LinInt,
$\max_{k=1}^m\{\rho(k)\}$ will be the Hausdorff distance between $P$ and LinInt,
$\delta_H(P,\text{LinInt})$, when $P$ is restricted to the region of
configuration space between the boundary conformations (and assuming that
structural alignment prior to rmsd measurement was performed
identically). Furthermore, when $P$ does not ``backtrack'', $\zeta(k)$ is
monotonically decreasing---indeed, $P$ can be said to backtrack (with respect to
some reference path) when $\zeta(k)$ is \emph{not} monotone---and the Hausdorff
and \frechet distances coincide: $\max_{k=1}^m\{\rho(k)\} =
\delta_F(P,\text{LinInt}) = \delta_H(P,\text{LinInt})$.

\paragraph{Heuristic collective variables.}
While dimensionality reduction can be useful for visualizing and identifying
functional protein motions, selecting the collective variables that span the
projected space and adequately describe a conformational transition is
nontrivial~\cite{Teodoro2003-lq, Mesentean2006-gl}. Choosing heuristic
coordinates for a given system often requires strong physical intuition,
something that may absent when studying new or complicated transitions. In
general, the determination of reaction coordinates and/or order parameters can
be guided by quantitative methods, such as principal component analysis or the
construction of isocommittor surfaces. In the relatively simple case of AdK's
\ctor transition, several viable order parameters have been used as
low-dimensional descriptions~\cite{Seyler2014-uu}.

To explicitly illustrate the uses and limitations of heuristic collective
variables, and to make a connection with previous work, we examine the AdK \ctor
transition (Fig.~\ref{fig:combotrans}A) in 2D angle-angle
space~\cite{Beckstein2009-ll}. The NMP-CORE angle \tnmp is formed by the
geometric centers of the backbone and C\textsubscript{$\beta$} atoms in residues
115-125 (CORE-LID), 90-100 (CORE), and 35-55 (NMP) of \textit{E.\ coli}
AdK. Likewise, \tlid is defined as the angle between residues 179-185 (CORE),
115-125 (CORE-hinge-LID), and 125-153 (LID). As many of the methods we
studied used \Ca-only models, we defined NMP-CORE and LID-CORE angles by
exclusively using the \Ca atoms of the residues. The angle-angle space defined by
(\tnmp, \tlid) quantifies the degree to which
NMP and LID are open and the sequence in which they open (close) for the \cto
(\otc) transition.

\subsection*{Generating transition paths}
\label{sec:pathwaygeneration}
We first describe the toy model system used to supply simple transitions for
testing purposes. We then summarize the path generation---using a variety of
enhanced path-sampling methods---of \cto transitions of AdK and DT, which
serve as more realistic representations of conformational transitions.

\paragraph{Toy model: Double-barrel potential.}
To determine the extent to which the Hausdorff and \frechet metrics are suitable
for measuring transition paths, we constructed a toy system to generate
well-defined trajectories driven by a one-way ramp potential and subject to
thermal noise; the resulting paths in configuration space can be viewed as
thermally-perturbed straight lines. For our purposes, transition progress was
measured by the center-of-mass distance of a group of particles moving along the
ramp so that a transition was completed once the center-of-mass trajectory
crossed a threshold value.

\begin{figure}[!htb]
\begin{adjustwidth}{0.5in}{0.5in}
  \centering
  \includegraphics[width=230pt]{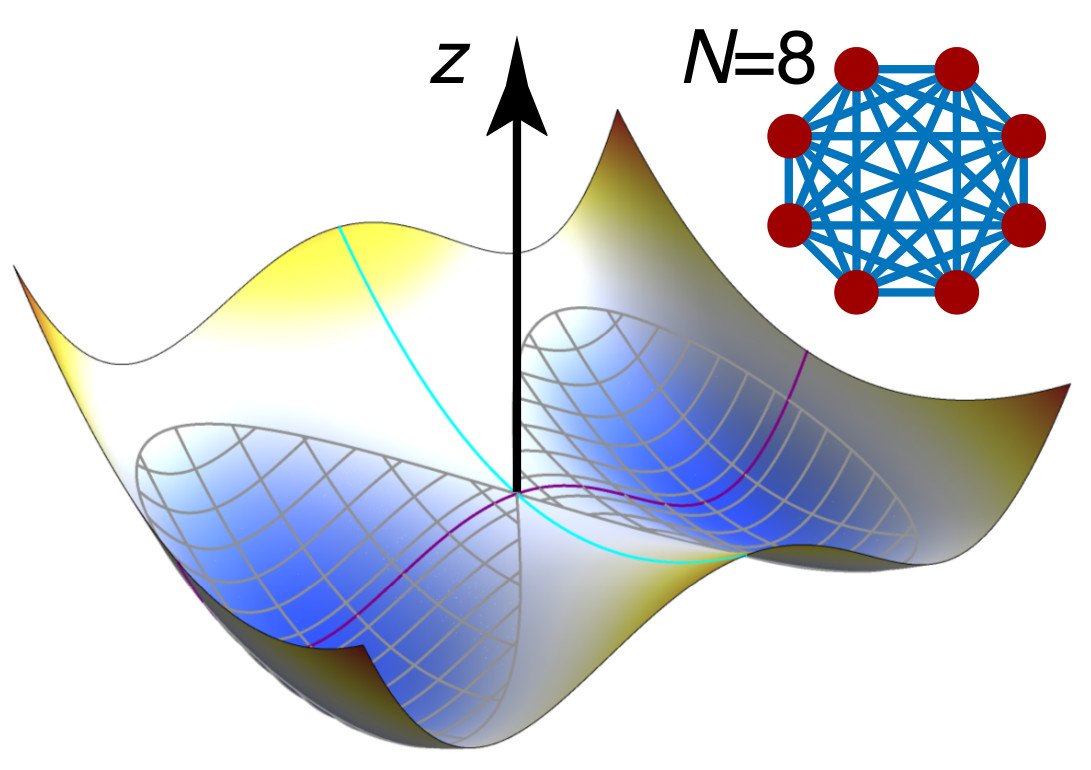}
  \caption{The toy model consists of a cluster of connected particles moving in
    a double-well potential along the $z$-axis under the influence of a linear
    ramp potential (not shown). In the cluster for $N=8$, each particle (red)
    is connected to every other particle with a harmonic spring (blue) of
    equilibrium length 0 (cluster not shown to scale.) The potential
    landscape for constant $z$ forms a ``double barrel''---red (blue)
    regions correspond to high (low) energies---is parabolic along the
    $x$-direction (cyan line), and has a double-well shape in the
    $y$-direction (purple line), which produces a  central barrier separating
    two ``barrels'' (gray crosshatching). A saddle point is located at the
    intersection of the cyan and purple lines. Motion in this landscape is
    biased toward either of the low-energy barrels, but transitions between
    barrels are possible at finite temperatures.}
  \label{fig:toy_model}
\end{adjustwidth}
\end{figure}

The toy system is defined as a group of $N$ particles connected by harmonic
springs subject to Brownian dynamics in a 3D potential energy landscape
(Figure~\ref{fig:toy_model}). Individual particles were connected in analogy to
a complete graph, with vertices and edges respectively representing particles
and springs. Spring equilibrium distances were set to zero separation for
simplicity. Differing dimensionalities of the configuration space were examined
by varying the number of particles $N$. The external potential was given a
double-well shape in the $y$-direction with a parabolic shape in the
$x$-direction (centered at $x=y=0$), ensuring that particle clusters are
confined to one of two ``barrels'' running along the $z$-direction
(Figure~\ref{fig:toy_model}). The energy barrier between the tubes was set to a
height of 2~$k_BT$ ($\sim\unit{5}{\kilo\joule\per\mole}$) at
$T=\unit{300}{\kelvin}$. We set up a ramp potential sloping down toward
increasing $z$ (i.e., a constant potential energy gradient in the positive $z$
direction) to induce large-scale transitions from small to large values of $z$.

To construct a properly coarse-grained system, we required zero-temperature
cluster dynamics to be identical for all $N$-particle clusters (given sensibly
chosen initial conditions). Spring constants, particle masses and sizes, and the
external potentials were scaled so as to preserve the average diffusive behavior
of a cluster. Furthermore, spring constants were chosen to be large enough to
prevent clusters from splitting themselves across the central barrier (where
some particles in the cluster fall to one tube and some fall to
other). Particles comprising a cluster were furthermore initialized at the same
location so that zero temperature center-of-mass trajectories would be
independent of particle number, $N$. It should be emphasized that this toy model
was not intended to replicate a real physical system, but primarily served to
build intuition prior to studying conformational pathways in realistic protein
systems. More detailed information about the construction of the double-barrel
system is provided in \nameref{S4_Text} in the Supporting Information.

\paragraph{Simulation methods and systems.}

The path-sampling methods comparison was performed using the AdK \cto transition
---between the (initial) closed conformation (PDB id 1AKE:A) and final (open)
conformation (PDB id 4AKE:A)---as a testbed. We used eight methods available on
publicly accessible servers \cite{Sfriso2012-uf, Sfriso2013-ah, Das2014-ry,
  Tekpinar2010-kc, Franklin2007-mz, Zheng2007-es, Krebs2000-di}, two in-house
methods (DIMS, dynamic importance sampling \cite{Perilla2011-im}, and FRODA,
Framework Rigidity Optimized Dynamics Algorithm \cite{Farrell2010-wh}), and
targeted MD (rTMD \cite{Ferrara2000-py}) using local simulation resources (see
Tables~\ref{tab:methods_energy} and \ref{tab:methods_path} for overviews). DIMS
and FRODA were additionally used to generate example ensembles of AdK and DT
transitions (200 transitions per method per protein, 800 total) for
ensemble-based and Hausdorff pairs analysis. In principle, other path-sampling
methods could be included in a comparison and, in the future, it would be worth
exploring alternative methods such as the finite-temperature string
method~\cite{Vanden-Eijnden:2009dq}, weighted ensemble
dynamics~\cite{Huber:1996ht, Zhang:2010ve},
milestoning~\cite{Bello-Rivas:2015vn}, transition path
sampling~\cite{Bolhuis:2002qe}, non-equilibrium umbrella sampling
\cite{Warmflash:2007kl} or forward flux sampling~\cite{Allen:2006cr}, to name a
few.  Key aspects of each method used in this study are summarized below to help
connect our results with physical intuition about the models. Each path-sampling
method is described in the context of the energetics they model
(Table~\ref{tab:methods_energy}) and the schemes by which paths are propagated
or generated (Table~\ref{tab:methods_path}). Additional details about the
methods and the corresponding simulation settings that were used can be found in
\nameref{S5_Text}.

\begin{table}[ht]
\begin{adjustwidth}{-2.25in}{0in} 
\setlength{\tabcolsep}{0.20em}
\caption{\sffamily\small\textbf{Modeling of energetics in tested path-sampling methods.}}
\sffamily\small
  \begin{tabular}{@{}|l|l|l|l|l|l|l|@{}}
    \hline
Res\tsup{a} & Name             & Force field/potential\tsup{b} & Solvent energetics\tsup{c} & Mixing function/other energetics\tsup{d} \\ \hline
all-atom
  & DIMS\cite{Perilla2011-im} & CHARMM22/CMAP & ACS/ACE2 IS          & $T=\unit{300}{\kelvin}$ \\\hline
  & rTMD\cite{Ferrara2000-py} & CHARMM22/CMAP & Generalized Born IS  & $T=\unit{300}{\kelvin}$ \\\hline
  & MDdMD\cite{Sfriso2012-uf} & bonds/angles: inf.\ sq-well & Lazaridis-Karplus IS  & NBF: simple vdW/electrostatic, $T=\unit{300}{\kelvin}$ \\\hline
  & FRODA\cite{Farrell2010-wh} & stereochemical constraints & hydrophobic contacts*  & overlap/angle/H-bond constraints \\\hline
  & Morph\cite{Krebs2000-di}  & CHARMM/XPLOR\tsup{$\dagger$} & -- & energy minimization of intermediate snapshots \\\hline
  & LinInt                    & -- & -- & -- \\\hline
\Ca-only
  & GOdMD\cite{Sfriso2013-ah}  & bonds: inf.\ sq-well      & -- & NBF: Go-like + ENM-MetaD \\\hline
  & ANMP\cite{Das2014-ry}      & double-well ANM         & -- & $E_\text{mix}=\min\left\{U_i,U_f\right\}$ \\\hline
  & iENM\cite{Tekpinar2010-kc} & double-well ANM & -- & $E_\text{mix}=F(U_i,U_f)$ (arbitrary), collision penalty \\\hline
  & MAP\cite{Franklin2007-mz}  & two ANMs, OM dynamics  & overdamped Langevin\tsup{$\ddagger$} & minimum OM action $\rightarrow$ 2 ODEs+BCs $\rightarrow$ path \\\hline
  & MENM-SD/SP\cite{Zheng2007-es} & double-well ANM         & -- & $E_\text{mix}=\beta^{-1} \ln\left\{\exp\left[-\beta(U_i+\epsilon_i)\right] + \exp\left(-\beta(U_f+\epsilon_f)\right]\right\}$ \\\hline
\end{tabular}
\begin{flushleft}
All MD-based methods use atomic resolution; Morph and LinInt are the only other methods
with greater than \Ca resolution. Except for MAP, ENM-based models define double-well potentials
using different mixing functions of each anisotropic network model (ANM) constructed about each native
states. MAP uses 2 ODEs, found by minimizing the Onsager-Machlup action
for each ANM about the native states, and satisfying continuity conditions for positions and
velocities at their interface. MENM-SD/SP assumes weak mixing: $T_m = T$ ($\beta=1/kT_m$, is an
adjustable parameter); in the limit of vanishing mixing, $T_m\rightarrow 0^+$,
$E_\text{mix}= \min\left\{U_i,U_f\right\}$, which is the same double-well potential used by ANMP.\\
\tsup{a}Resolution of the model. \\
\tsup{b}inf. sq-well, infinite square well; ANM, anisotropic network model; OM, Onsager-Machlup. \\
\tsup{c}IS, implicit solvent; FRODA does not have a solvent model; MAP assumes overdamped
Langevin dynamics in using the Onsager-Machlup action. \\
\tsup{d}NBF, non-bonded forces; vdW, van der Waals potential; ENM-MetaD, elastic network model-based metadynamics; $E_\text{mix}$,
mixing function for two-state potential; $U_i$ ($U_f$), potential energy function about the initial
(final) native state; OM, Onsager-Machlup; ODEs+BCs, ordinary differential equations plus boundary conditions. \\
\tsup{*}FRODA does not use a solvent model. \\
\tsup{$\dagger$}Morph uses CHARMM/XPLOR relaxation to minimize energy of intermediate snapshots. \\
\tsup{$\ddagger$}MAP assumes overdamped Langevin dynamics in using the Onsager-Machlup action. \\
\end{flushleft}
\label{tab:methods_energy}
\end{adjustwidth}
\end{table}

\begin{table}[ht]
\begin{adjustwidth}{-2.25in}{0in} 
\setlength{\tabcolsep}{0.20em}
\caption{\sffamily\small\textbf{Approach to generating paths in tested path-sampling methods.}}
\sffamily\small
  \begin{tabular}{@{}|l|l|l|l|l|l|l|@{}}
    \hline
Type       & Name                      & Dynamics & Path propagation/biasing\tsup{a} & Rev\tsup{b} & TS/Stoch\tsup{c} & Progress variable \\ \hline
perturbation MD
  & DIMS\cite{Perilla2011-im} & Langevin NVT   & SR                        & N & Y/Y & rmsd-to-target \\\hline
  & rTMD\cite{Ferrara2000-py} & Langevin NVT   & moving harmonic restraint & N & Y/Y & rmsd-to-target \\\hline
  & MDdMD\cite{Sfriso2012-uf} & discrete MD    & SR + essential dynamics   & N & Y/Y & ssd-to-target\tsup{$\dagger$} \\\hline
  & GOdMD\cite{Sfriso2013-ah} & discrete CG-MD & SR + metadynamics         & N & Y/Y & ssd-to-target\tsup{$\dagger$} \\\hline
geometric targeting
  & FRODA\cite{Farrell2010-wh} & -- & stepwise-enforced rmsd constraint*  & N & Y/(Y/N) & rmsd-to-target \\\hline
CG-ENM
  & ANMP\cite{Das2014-ry}                    & -- & SD from SP (cusp min.) to minima & Y & N/N & -- \\\hline
  & iENM\cite{Tekpinar2010-kc}               & -- & parametric SP/fixed-point eqn.   & Y & N/N & -- \\\hline
  & MAP\cite{Franklin2007-mz}                & -- & OM minimum action path           & Y & N/N & -- \\\hline
  & MENM-SD\cite{Zheng2007-es}               & -- & SD from SP to minima             & Y & N/N & -- \\\hline
  & MENM-SP\cite{Zheng2007-es}               & -- & parametric SP/fixed-point eqn.   & Y & N/N & -- \\\hline
adiabatic mapping & Morph\cite{Krebs2000-di} & -- & linearly interpolated snapshots  & Y & N/N & -- \\\hline
linear interpolation & LinInt                & -- & linearly interpolated snapshots  & Y & N/N & -- \\\hline
\end{tabular}
\begin{flushleft}
DIMS, rTMD, MDdMD, and GOdMD are all non-deterministic MD-based methods.
DIMS and rTMD employ a conventional force field and Langevin dynamics in the canonical
ensemble; the discrete MD algorithms used by MDdMD and GOdMD assume ballistic particle
motion until a collision occurs---along with the depth of the interatomic square wells,
momentum and energy conservation are used to determine outgoing momenta without explicitly
computing forces. FRODA uses a non-physical dynamical algorithm to path-search
stereochemically correct regions of configuration space. CG-ENM methods generate
transitions by constructing low-energy paths in the potential energy landscape. Morph
and LinInt linearly interpolate the position of each atom between the initial and final
states. \\
\tsup{a}SR, soft ratcheting; SD, steepest descent; SP, saddle point; OM, Onsager-Machlup. \\
\tsup{b}Is the method exactly reversible? \\
\tsup{c}Is the algorithm based on a (physical or non-physical) time step? Is it stochastic? \\
\tsup{*}At each step, rmsd reduced by fixed amount while simultaneously enforcing other constraints. \\
\tsup{$\dagger$}ssd, sum of squared distances to target (includes weighting that varies
between MDdMD and GOdMD).
\end{flushleft}
\label{tab:methods_path}
\end{adjustwidth}
\end{table}

Two of the tested methods are based on MD combined with perturbation techniques
(perturbation MD) designed to drive transitions between initial and final
states. Two MD methods---\textbf{DIMS} MD~\cite{Woolf1998-xh, Perilla2011-im}
(implemented in CHARMM c36b2 \cite{Brooks2009-be}) and TMD (implemented in NAMD
2.10~\cite{Phillips2005-hx})---used the all-atom CHARMM22/CMAP force field
\cite{MacKerell98, MacKerell04a} with Langevin dynamics and implicit solvents
(ACE~\cite{Schaefer:2001et} in CHARMM, Generalized Born~\cite{Tanner:2011oq} in
NAMD) in the NVT ensemble at \unit{300}{\kelvin}. NAMD's TMD implementation uses
a time-dependent harmonic restraint that moves toward a target conformation with
constant velocity~\cite{Ferrara2000-py}, instead of the original TMD approach
introduced by Schlitter~\cite{Schlitter1994-xz} that employed a stepwise
holonomic constraint; in accordance with Ovchinnikov and
Karplus~\cite{Ovchinnikov2012-pf}, we refer to the NAMD implementation as
\emph{restrained TMD} (\textbf{rTMD}) to distinguish it from the original
algorithm.  DIMS and rTMD transitions were driven using the heavy-atom rmsd to
the target structure. The soft-ratcheting DIMS algorithm moves towards the
target by taking trial MD steps. Steps toward the target (decreasing
rmsd-to-target) are accepted whereas backward steps are rejected with a finite
probability; velocities are re-sampled (according to Maxwell-Boltzmann) until
the step is accepted. rTMD moves a harmonic restraint to linearly decrease the
rmsd-to-target. We generated three rTMD paths using a fast pulling speed and
three with slower pulling. rTMD differs from DIMS in that explicit forces are
introduced into the system Hamiltonian whereas DIMS effectively introduces an
entropic force.

Maxwell-Demon discrete Molecular Dynamics~\cite{Sfriso2012-uf} (\textbf{MDdMD})
and \textbf{GOdMD}~\cite{Sfriso2013-ah} are similar in spirit and are the only
two methods based on a physical dynamical model among the server-based
transition path generation methods. Both are based on discrete MD combined with
soft ratcheting and a type of essential dynamics sampling. MDdMD utilizes an
atomistic representation and an implicit solvent model; GOdMD bonds uses a \Ca
representation and neglects solvent effects. Both methods model bonded forces
with infinite square-wells although MDdMD incorporates further detail by using
simple potentials to describe van der Waals and electrostatic forces; GOdMD uses
a Go-like potential to describe non-bonded forces. Both also include an
additional form of biasing to ensure transitions follow essential deformation
movements of a protein: MDdMD accepts steps when the transition vector (from the
current conformer to the target) overlaps sufficiently with an essential
transition vector (defined using eigenvectors from NMA on a Go-like potential
about the initial or final state); GOdMD uses an ENM-based metadynamics approach
to bias the sampling of essential deformation modes and to ensure that
trajectories escape the initial minima.

The geometrical targeting algorithm, \textbf{FRODA}~\cite{Farrell2010-wh}, is an
approach designed to produce stereochemically acceptable transition paths. FRODA
moves a structure toward a target conformation by decreasing the rmsd-to-target
while enforcing stereochemical constraints such as bond distances and angles,
backbone dihedrals, and contact constraints. In particular, FRODA can avoid
steric clashes in an all-atom structure, something that may not be achieved by
coarse-grained elastic network models (CG-ENMs) or algorithms using simple linear
interpolation.

We also generated transitions using five CG-ENM-based methods. These
particular models first construct two harmonic potential energy functions, based on
anisotropic network models (ANMs), about initial and
final native (crystallographic) states, which has the general form
\begin{displaymath}
  \label{eq:anm}
  U(\mv{X}) = \frac{1}{2}\sum_{d_{ij}^{\,0}<R_c} 
              C_{ij}\left(d_{ij}-d_{ij}^{\,0}\right)^2 + \Delta U,
\end{displaymath}
where the sum is taken over all unique pairs of \Ca atoms separated by less than
a specified cutoff distance, $R_c$, and $\Delta U$ is the energy difference
between the two states. For atoms $i$ and $j$, $C_{ij}$ is the force constant,
$d_{ij}$ is the Euclidean distance between them, and $d_{ij}^{\,0}$ is the
corresponding distance in the native (crystallographic) structure. Force
constants can determined by fitting to isotropic crystallographic B-factors for
instance. A double-well (two-state) potential landscape is constructed by
combining the separate potentials. Given a two-state potential, transition paths
are generated by connecting the two (end-state) minima along low-energy
pathways. The ENM-based methods are distinguished primarily by their two-state
energetics (i.e., mixing potential) and method of defining and searching for
low-energy transition paths. The cutoff distance, $R_c$, can adjusted to some
degree for all the tested ENM-based approaches, but a couple also enable
modification of the force (spring) constants, $C_{ij}$, and the end state energy
difference, $\Delta U$.

ANMPathway~\cite{Das2014-ry} (\textbf{ANMP}) forms a double-well landscape by
taking the lower of the individual wells; the wells intersect to form a cusp
hypersurface in configuration space. A path is found by locating the minimum
along the cusp and performing steepest descent (SD) toward both well minima. The
Mixed Elastic Network Model~\cite{Zheng2007-es} (\textbf{MENM}) employs a
double-well function with a tunable mixing temperature whose purpose is to
module the cusp-like intersection to provide a smooth transition in energy
between the two wells. The method locates saddle points (SPs), and can use a
steepest descent (SD) mode (\textbf{MENM-SD}) from the SPs to the minima, or it
can provide a parametric equation describing an SP path through the fixed points
of the landscape. The interpolated Elastic Network Model~\cite{Tekpinar2010-kc}
(\textbf{iENM}) is similar to MENM-SP in that it analytically solves for a
parametric SP path, although it only requires a general form of a double-well
potential function (does not use an explicit mixing function). Unlike the other
ENM-based methods, MinActionPath~\cite{Franklin2007-mz} (\textbf{MAP}) does not
use a mixing function. Instead, a path is generated by minimizing the Onsager-
Machlup (OM) action---which assumes overdamped Langevin dynamics---with the two
separate ANMs for the native states to derive two one-dimensional differential
equations describing the minimum action paths in the region of each ANM. A
unique transition path between the initial and final states is found by
satisfying continuity boundary conditions in the positions and velocities at the
interface.

To provide a point of comparison to one of the most simple approaches, we used
the Yale morph server~\cite{Krebs2000-di, Flores2006-bf} (\textbf{Morph}), which
combines linear interpolation and optional energy minimization of the
intermediate snapshots (i.e., adiabatic mapping), and we also used explicit
linear interpolation to generate a single path between the end states
(\textbf{LinInt}).

The main thrust of the path-sampling methods comparison is to demonstrate PSA's
viability and not necessarily to directly evaluate the performance of the
sampling algorithms. As such, adjustable parameters for all simulations were
left at their default values unless explicitly stated.  Transitions were
produced using the highest allowable resolution, i.e., using all non-hydrogen
atoms when possible or only \Ca atoms otherwise. For each method, three unique
paths were generated by either re-running those with stochastic algorithms or,
for the deterministic ones, by adjusting a single parameter; in the case
  of rTMD, six total simulations were performed [three each for fast
  ($\sim\unit{1}{\angstrom\per\pico\second}$) and slow
  ($\sim\unit{0.01}{\angstrom\per\pico\second}$) pulling speed; see
  \nameref{S5_Text} for further details]. DIMS, FRODA and MDdMD simulations,
which produce a unique trajectory every run, were run three times each without
altering initial settings. Three GOdMD runs were performed by changing the
relaxation window ($\unit{20}{\pico\second}$, $\unit{50}{\pico\second}$ and
$\unit{100}{\pico\second}$). Distinct trajectories for the deterministic, ENM-
based algorithms were obtained by varying spring cutoff distances: one
transition at the default value and two by decreasing/increasing the cutoff.
Morph trajectories were produced by toggling energy minimization and structural
pre-alignment settings, and a single LinInt trajectory was included as a
zeroth-order reference. All other simulation settings were left at default
values where possible. Simulations and analyses performed in this study are
summarized in Table~\ref{tab:tests}. Furthermore, as half of the methods were
limited to \Ca structures as inputs---the coarsest representation among the
methods---all analyses were restricted to \Ca trajectory representations to
provide a lowest common denominator. Trajectories were also aligned to a common
reference structure generated by aligning and averaging the CORE \Ca coordinates
of the 1AKE:A and 4AKE:A structures (see \nameref{S6_Text} in the Supporting
Information for a description of the structural alignment procedures).

\begin{table}[ht]
\begin{adjustwidth}{-2.25in}{0in} 
\caption{\sffamily\small\textbf{Summary of simulations, calculations, and analyses.}}
\sffamily\small
\begin{tabular}{@{}|l|l|l|l|l|l|@{}}
\hline
Assessment                 & System & Transition                 & Path generation
      & \# path samples    & Analysis methods\tsup{$\dagger$} \\
\hline
(1) Intuition and viability        & double-barrel & $z$:~$0\rightarrow 4$~nm & Brownian$\,+\,$ramp
      & 4$\times$(2 ICs)   & PSA (\dF), \dF{}-\dH distr/corr*            \\ \hline
(2) Methods comparison     & AdK & \cto   & various methods
      & 3$\times$(11 methods)  & PSA (\dF/\dH{}*), NCA, \zr, AA   \\ \hline
(3) Transition ensembles  & AdK & \cto  & DIMS, FRODA & 200$\times$(2 methods) & PSA (\dF{}*), \dF-\dH distr/corr*    \\ \hline
              & DT & \cto  & DIMS, FRODA & 200$\times$(2 methods) & PSA (\dF), \dF-\dH distr/corr*  \\ \hline
(4) Atomic detail from PSA & AdK & \cto    & DIMS, FRODA
      & 200$\times$(2 methods) & PSA (\dH-pairs)   \\
\hline
\end{tabular}
\begin{flushleft}
*Result in Supporting Information \\
\tsup{$\dagger$}Analysis methods: PSA, path similarity analysis;
\dF, \frechet distance; \dH, Hausdorff distance; \dF-\dH distr/corr, \frechet{}-Hausdorff
distribution/correlation analysis; NCA, native contacts analysis;
\zr, progress vs. displacement along path of linear interpolation; AA, angle-angle space.
\end{flushleft}
\label{tab:tests}
\end{adjustwidth}
\end{table}

\section*{Results and Discussion}
We subdivided our study in four parts to show how PSA can be used to answer a
range of questions about macromolecular transition paths and pathways (see
Table~\ref{tab:tests}):
(1) The path metrics were able to distinguish and categorize simple trajectories
in a toy system, taking thermal motion and varying number of particles into
account. (2) PSA could be used to compare different path-sampling methods and,
when combined with more traditional low-dimensional projections on collective
variables, provide insights into similarities and differences between different
methods. (3) PSA was able to analyze path ensembles, opening the door
to analyzing dynamical simulations with statistical approaches. (4) PSA
enabled us to extract the molecular structural determinants responsible for
differences in paths, thus linking the general analysis of high-dimensional
transition paths to the specific molecular detail.

\subsection*{Path similarity analysis of toy model transitions}
\begin{figure}[!htb]
\begin{adjustwidth}{-2.0in}{0.5in}
  \centering
  \includegraphics[]{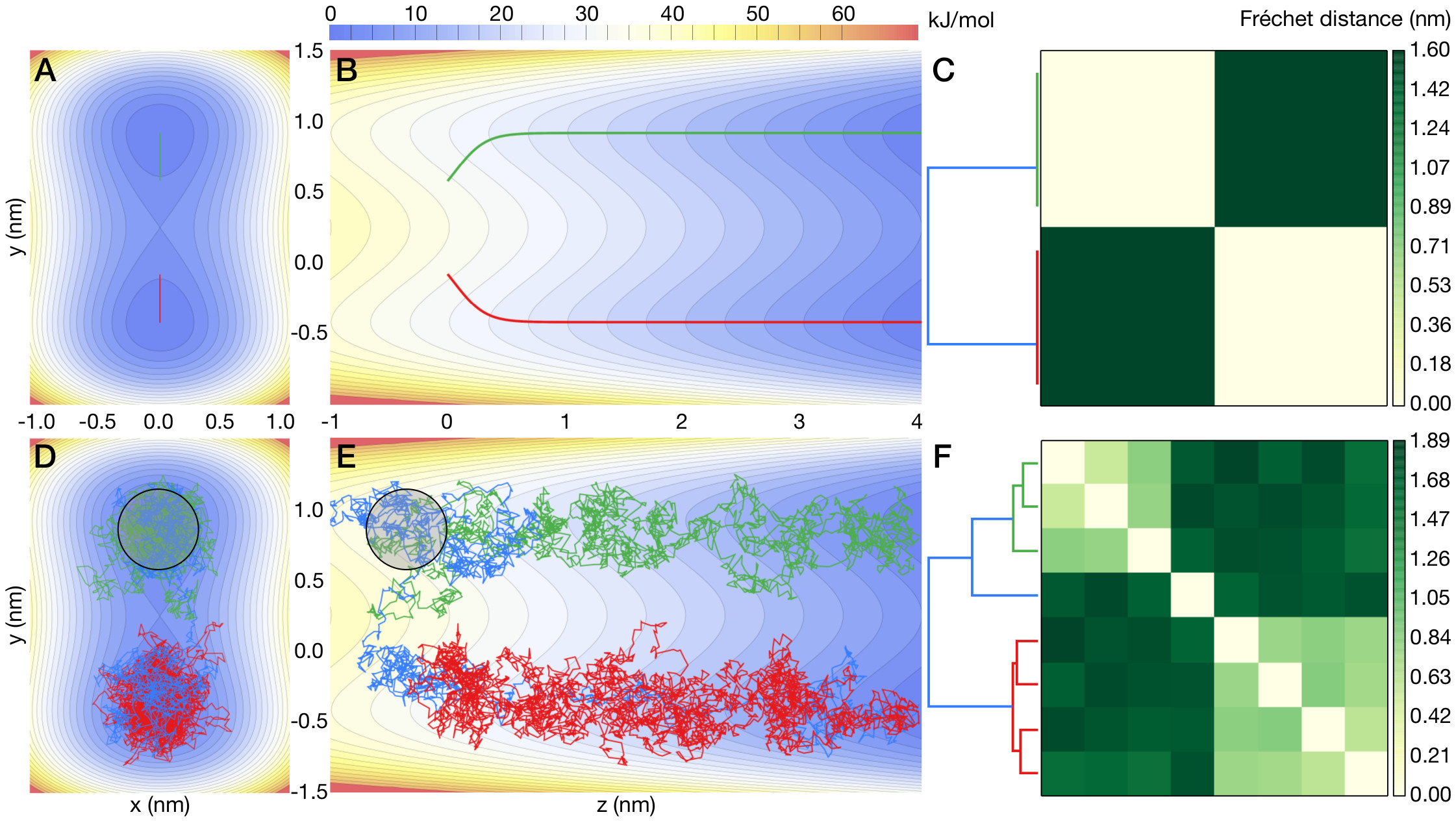}
  \caption{Double-barrel potential energy landscape projected
    onto the $xy$-plane and $yz$-plane. Groups of point masses (clusters) mutually
    connected by harmonic springs move under the influence of a transition-inducing
    ramp potential in the positive $z$ direction and the two low-energy minima
    of the ``barrels'' at $y=\pm \unit{0.8}{\nano\meter}$. Colored lines depict
    the center of mass trajectories for each cluster. (A--C) trajectories at
    $T=\unit{0}{\kelvin}$. (D--F) trajectories at $T=\unit{250}{\kelvin}$.
    (A, D) Projection of paths onto the $xy$-plane together with the double-barrel
    potential. (B, E) Projection of paths onto the $yz$-plane. (C, E) Clustered
    heat maps summarize the \frechet distances for all pairs of trajectories;
    dendrograms record cluster distances according to the Ward criterion.
    Trajectory colors in each row match the corresponding path(s) in the
    dendrogram. The trajectory-averaged radius of gyration for clusters at
    finite temperature is \unit{0.35}{\nano\meter} (black circles).
    }
  \label{fig:doublebarrel}
\end{adjustwidth}
\end{figure}

We simulated one- and eight-particle cluster transitions in the double-barrel
potential energy landscape between a starting state (defined as a center-of-mass
location below $z=\unit{0}{nm}$) and a final state
($z\geq\unit{4}{nm}$). Eight-particle simulations at zero and
\unit{250}{\kelvin} are shown in Fig.~\ref{fig:toy_model}. The particles were
weakly confined to one of two potential energy barrels separated by a 2~$k_BT$
barrier at \unit{300}{\kelvin} (Fig.~\ref{fig:doublebarrel}A,D) and evolved
under the influence of thermal diffusion and drift due to a linearly decreasing
ramp potential in the $z$ direction
(Fig.~\ref{fig:doublebarrel}B,E). Simulations were run at temperatures between
\unit{0}{\kelvin} and \unit{600}{\kelvin} in \unit{50}{\kelvin} increments, with
eight runs at each temperature. Trajectories were initialized such that two
distinct groups of paths would be produced at zero temperature: for each
temperature, we initialized half of the simulations to one side of the central
barrier at $(x_0,y_0) = (\unit{0}{nm},\unit{0.4}{nm})$ and the other half at
$(\unit{0}{nm},\unit{-0.4}{nm})$.

At zero temperature, trajectories initiated at the same point progressed along
identical paths due to the absence of thermal diffusion. Two trajectory groups
were formed (Fig.~\ref{fig:doublebarrel}A,B), consistent with what was
expected from the initial conditions. A clustered heat map of the \frechet
distances between the $T=\unit{0}{\kelvin}$ trajectories clearly showed two
well-defined clusters (Fig.~\ref{fig:doublebarrel}C), containing four
trajectories each, in both the structure of the dendrogram as well as the color
division in the heat map. Due to thermal perturbations, higher-temperature
trajectories exhibited substantial wandering (Fig.~\ref{fig:doublebarrel}D,E)
and even produced a transition across the central barrier (blue trajectory in
Fig.~\ref{fig:doublebarrel}E). In contrast with the zero temperature case,
both the number of clusters and the clusters themselves were much more vaguely
defined. Two clusters with four trajectories per cluster (red and green/blue
trajectories, Fig.~\ref{fig:doublebarrel}D--F) were still formed, although
the blue trajectory, which underwent a barrier-crossing transition near
$z=-\unit{0.5}{\nano\meter}$, is an outlier in the cluster with the three
green trajectories.

Trajectory categorization for the toy model with PSA did not depend strongly on
the dimensionality (cluster size) as thermal noise alone appeared to have a much
more substantial influence (\nameref{S1_Fig}). In particular, we could not
discern meaningful differences in the center of mass motions between one- and
eight-particle clusters from the data. Furthermore, in the eight-particle case
at \unit{250}{\kelvin}, performing PSA using the full ($24$-dimensional)
configuration space trajectories did not produce a different clustering than PSA
applied only to the center of mass trajectories (data not shown).  The same
analysis as above was carried out with the Hausdorff distance instead of the
\frechet distance to assess their relative discriminative powers. Both produce
similar results at temperatures below \unit{300}{\kelvin} with low-temperature
simulations exhibiting two distinct pathways (\nameref{S2_Fig}). Between
\unit{350}{\kelvin} and \unit{500}{\kelvin}, however, Hausdorff and \frechet
distance measurements started to become substantially uncorrelated
(\nameref{S3_Fig}). This effect is likely due in part to the sensitivity of the
\frechet metric to backtracking (Fig.~\ref{fig:frechethaus}), which may be
amplified when the typical energy of thermal perturbations become comparable to
the height of a potential barrier ($2 k_{B}T$ at
\unit{300}{\kelvin}). High-temperature simulations ($\geq$\unit{300}{\kelvin})
began to explore both tubes as if they were a single pathway (\nameref{S2_Fig}
and \nameref{S4_Fig}).

Taken together, PSA was able to distinguish groups of paths in the presence of
stochastic thermal motions as long as the thermal energy was lower than the
energy scale of distinguishing features in the underlying energy landscape. The
dimensionality of the problem did not appear to be an important factor.
\frechet and Hausdorff distances discriminated paths equally well with some
small differences at high temperatures that likely reflect backtracking of
trajectories.

\subsection*{Comparing enhanced path-sampling methods}
In order to compare a selection of fast transition path sampling methods, three
distinct trajectories were generated for the \cto AdK transition as described in
Methods.

\begin{figure}[htb]
\begin{adjustwidth}{-0.5in}{0.5in}
  \centering
  \includegraphics[]{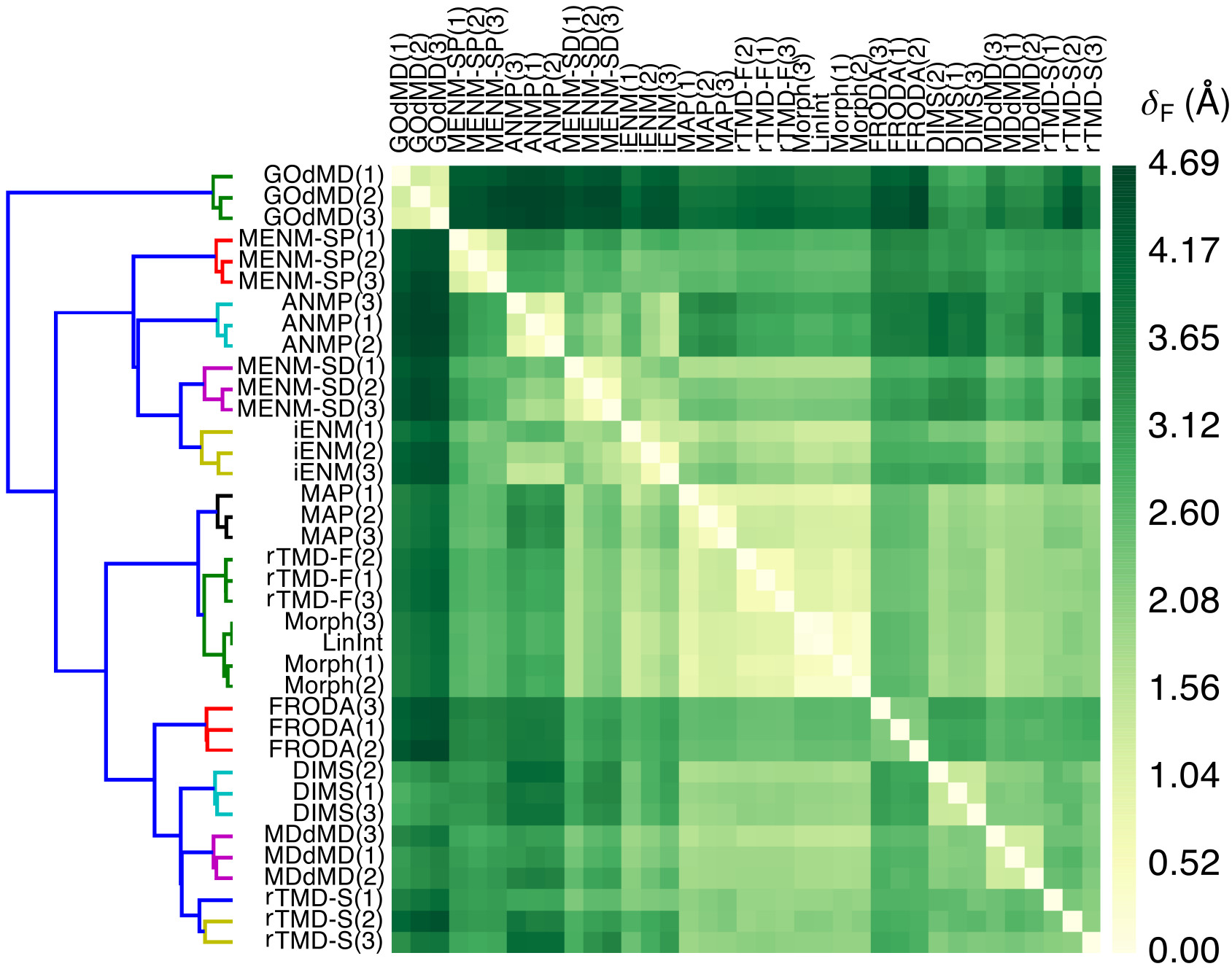}
  \caption{Path similarity analysis of trajectories generated by different
    path-sampling methods. The AdK closed$\rightarrow$open transition was
    sampled three times (except LinInt) with different methods (see
    text). Smaller distances indicate transition paths with greater
    similarity. The dendrogram depicts a hierarchy of clusters where smaller
    node heights of parent clusters indicate greater similarity between child
    clusters. \frechet distances $\delta_{F}$ are in \AA{} and correspond to a
    structural rmsd in accordance with the rmsd point metric. See text for a
    description of the methods. \protect\nameref{S5_Fig} contains the
      same data annotated with numerical values of $\delta_{F}$.}
  \label{fig:m_psa}
\end{adjustwidth}
\end{figure}

\paragraph{Direct comparison using PSA.}
A total of 37 paths (eleven methods, three paths
per method with the exception of six paths for rTMD, plus one LinInt
path) were analyzed by computing the \frechet distance between all possible
pairs and clustering of the resulting (symmetric) distance matrix
(\nameref{S5_Fig}). Using the same approach as with the toy model, the clustered
distance matrix was translated to a heat map-dendrogram representation
(Fig.~\ref{fig:m_psa}).

Paths from a given method were more similar to other paths from the same method
than to those produced by a different method, as indicated by well-defined
$3\times3$ squares along the heat map diagonal. Methods based on similar
physical models tended to produce relatively similar pathways, while
algorithmically distinct approaches appeared less likely to produce similar
pathways. For instance, Morph and LinInt both implement linear coordinate
interpolation. Their paths are essentially identical ($\delta_{F} \leq
0.5$~\AA), which indicates that additional features implemented in Morph, such
as checking for steric overlaps, may not be relevant for the AdK
transition. Another cluster was formed by the two MD-based importance sampling
methods, DIMS and MDdMD, together with MD-based rTMD at slow pulling
  velocity (``rTMD-S''; \frechet distance $2.1\ \mathrm{\AA} \leq \delta_{F}
  \leq 2.7\ \mathrm{\AA}$). In other cases, similarities and differences did
not always follow an immediately obvious pattern. FRODA, which satisfies
rigidity constraints during a transition but does not employ a potential energy
function, nevertheless formed a cluster with DIMS, MDdMD, and rTMD-S
($2.6\ \mathrm{\AA} \leq \delta_{F} \leq 3.1\ \mathrm{\AA}$). The grouping of
FRODA with DIMS/MDdMD/rTMD-S appears, however, less strong than, for
instance, the clustering of DIMS with MDdMD because for other choices of the
linkage algorithm FRODA is more distantly associated with the
DIMS/MDdMD/rTMD-S cluster and a robust cluster of MAP/Morph/LinInt
trajectories (see \nameref{S6_Fig} B--D and further discussion in
\nameref{S3_Text}). The fast-pulling rTMD (``rTMD-F'') and MAP
trajectories were strikingly similar to the Morph paths ($\delta_{F} \approx
1$~\AA), even though rTMD-F performs MD with an atomistic physics-based
  force field, whereas MAP's energy function is based on an elastic network
model and the path is generated via minimization of Onsager-Machlup action (and
not just linear interpolation). Interestingly, the MAP/rTMD-F/Morph
sub-cluster was grouped with the cluster formed by four of the
dynamical algorithms (DIMS, MDdMD, rTMD-S, FRODA). The other four ENM
algorithms---iENM, MENM-SD/SP, and ANMP---produced their own cluster, with
MENM-SD and iENM being the most similar to each other. A careful examination of
the heat map revealed that although MAP, rTMD-F, and Morph paths somewhat
resembled iENM and MENM-SD paths ($\delta_{F} \leq 2.5$~\AA), their overall
patterns of \frechet distances were very similar to DIMS/MDdMD/rTMD-S
(as seen in the similar overall striping in the shading of the heat map)
so that the ``Morph-like cluster'' rather clustered with these dynamical
  methods than with the ``ENM cluster''. The GOdMD paths formed their own
outlier cluster, appearing substantially different from all other methods
($\delta_{F} > 3$~\AA).

The classification of trajectories was found to be robust against use of
different linkage functions in the clustering algorithm, provided that the
linkage primarily assessed the \emph{dissimilarity} of clusters (such as Ward's
criterion in Fig.~\ref{fig:m_psa} and the complete/average/weighted linkage in
\nameref{S6_Fig} B--D) instead of similarity (single linkage in \nameref{S6_Fig}
A). Using the Hausdorff metric instead of the \frechet
metric did not change the clustering either and the Pearson
correlation coefficient between $\delta_{H}$ and $\delta_{F}$ was very close to
unity (\nameref{S7_Fig}). In \nameref{S1_Text}, alternative distance
  definitions, namely averaged \frechet and Hausdorff distances (which are,
  however, not proper metrics), reduced the amount of detail in the clustering
  and resulted in an amalgamation of clusters into one large ``dynamical
  methods cluster'' (TMD-S, DIMS, MDdMD, GOdMD, FRODA), a ``Morph-like cluster''
  (Morph, LinInt, TMD-S, MAP), and an ``ENM cluster'' (ANMP, iENM, MENM-SP/SD).

Without any input except the trajectories themselves, PSA produced distinct
clusters that appeared to broadly distinguish between dynamical and
non-dynamical path sampling methods. With the help of more specialized analyses
to be described next we sought to further rationalize the observed pattern of
clustering.

\begin{figure}[]
\begin{adjustwidth}{-1.0in}{0.5in}
  \centering
  \includegraphics[]{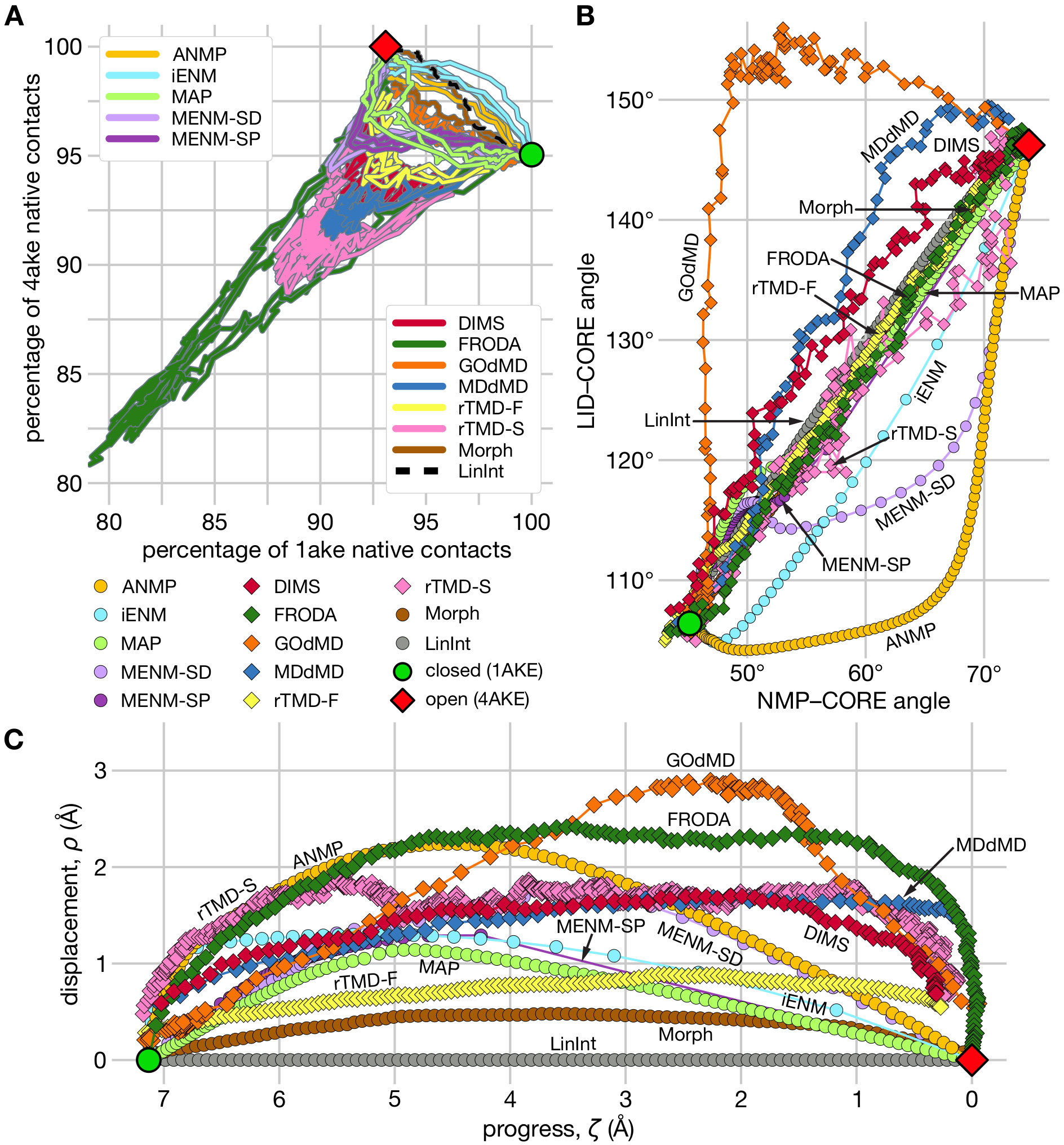}
  \caption{Projections of trajectory 2 of the AdK \cto transition from each
    path-sampling method onto low-dimensional collective variables. The location
    of the initial structure is shown in each plot by the green circle, while
    the final structure is represented by the red diamond.  (A) Projection of
    all pathways from the various path-sampling methods onto NC space. The
    horizontal axis corresponds to the percentage of contacts (of a transition
    snapshot) shared with the initial 1AKE:A structure (green circle) and the
    percentage of contacts in common with the final 4AKE:A structure (red
    diamond) is displayed on the vertical axis. The top-left legend identifies
    EN-based methods; the other methods are listed in the bottom legend.  The
    LinInt path is shown for reference as a broken black curve. (B) Projection
    on NMP angle ($\theta_{\text{NMP}}$) \emph{vs} LID angle
    ($\theta_{\text{LID}}$). In B and C, trajectories generated by the dynamical
    methods (DIMS, rTMD, FRODA, MDdMD, GOdMD) are plotted with diamonds
    and non-dynamical method trajectories with circles. (C) \zr space projection
    using LinInt as the reference path. Trajectory progress in \zr space is from
    left to right from higher to lower values of the progress variable
    $\zeta$. MDdMD terminates at 1.5~\AA~\Ca rmsd from 4AKE (red diamond); DIMS
    MD terminates at 0.5~\AA~heavy atom rmsd.}
  \label{fig:m_cv}
\end{adjustwidth}
\end{figure}

\paragraph{Native Contacts Analysis.}
We performed two dimensional NCA on trajectories by measuring (for each
conformer snapshot) the fraction of native contacts relative to the closed
starting state (\qc) and to the open target conformation (\qo) as collective
variables (Fig.~\ref{fig:m_cv}A). Using the NC trajectories, we examined the
dynamic relationship of contact formation and breaking for each method. In
general, the \cto trajectories began on or near the right vertical axis,
corresponding to the first conformers of the paths having (nearly) 100\% of
their contacts in common with the closed structure and around 95\% of open
state contacts. Most trajectories terminated at the top horizontal axis with
the final conformers containing close to 100\% of the final, open 4AKE:A
structure contacts and about 93\% of 1AKE:A contacts. The starting conformers
of the DIMS NC paths only contained 96\% of the contacts seen in the 1AKE
crystal structure (\qc~$=0.96$), which is to be expected given that the initial
closed structure was energy-minimized and equilibrated prior to performing
MD.

The five dynamical methods---DIMS, rTMD, FRODA, MDdMD,
and GOdMD---produced somewhat noisy paths where the fluctuations took place
along a positively sloping direction in NC space. A positive slope implies that
contacts were simultaneously formed or broken relative to both native
structures, which can be taken to be indicative of passage through a transition
state that is distinct from either end state conformation. DIMS trajectories did
not exactly reach the target structure (\qo~$\leq 0.98$) as DIMS simulations
were considered complete as soon as a conformer was within 0.5~\AA{} heavy atom
(non-hydrogen) rmsd from the target crystal structure 4AKE. MDdMD paths partly
overlapped with DIMS paths during contact breaking but failed to reform them
(\qo~$<0.94$); as with DIMS, transition completion is determined by a
cutoff---manually set to 1.5~\AA~\Ca rmsd---due to the difficulties of
convergence to a target using the soft-ratcheting biasing approach in
MDdMD. DIMS and MDdMD broke a similar number of contacts relative to both states
(around 8-9\% and 9-10\%, respectively). rTMD-S showed qualitatively
  similar behavior but broke up to about 12\% of native contacts. The
closely-knit cluster of DIMS, MDdMD and rTMD-S paths produced by PSA reflects the
qualitative similarity of their NC trajectories. DIMS, MDdMD and FRODA all
generated noisy, V-shaped NC pathways suggestive of a transition region and
supports the picture from PSA where these three methods form a loose cluster
apart from the non-dynamical methods. FRODA clustered somewhat apart from the
other three, which correlates with the observation that FRODA
trajectories in NC space exhibited the greatest contact breaking (\qc~$=0.82$,
\qo~$=0.80$) of all methods tested. This behavior is not unexpected because
FRODA achieves random motion by randomly displacing and rotating rigid units of
the protein at the sub-amino acid level at each step prior to re-enforcing
geometric constraints. As such, \Ca fluctuations and, thus, native contact
dynamics that would be prohibited by conventional potentials are permitted by
the geometric model although constraints on the overall sequence and structure
would nevertheless limit dramatic perturbations to the \Ca rmsd. GOdMD paths,
though quite noisy, followed a path more closely resembling those from the
non-dynamical methods, particularly MAP and Morph.

Morph, LinInt, and two of the five ENM-based methods (ANMP and iENM) produced
the shortest NC trajectories progressing directly to the target conformation
with relatively little wandering, whereas the six MENM paths deviated noticeably
toward the DIMS and FRODA trajectories in the latter half of the transition; MAP
paths were also nearer the MENM pathways in location and shape than to the paths
from the other ENM-based methods. The MENM paths and two MAP paths were unique
among the non-dynamical methods in that they each contained a V-shaped,
cusp-like feature where extra 4AKE:A contacts were broken (\qo~$\approx 0.91$,
$0.91$ and $0.92$, respectively) that were subsequently reformed toward the end
of the transition. The rTMD-F NC path was situated in an intermediate
  position between the other dynamical methods and the non-dynamical
  methods. Initially, only 1AKE:A contacts that do not exist in 4AKE:A were
  broken. Then the missing 4AKE:A native contacts were formed. The Morph,
LinInt, ANMP and iENM paths, which were divided between two clusters in PSA,
exhibited progress along negatively sloped NC space trajectories during which
4AKE:A contacts were formed while 1AKE:A contacts were simultaneously
broken. However, the close structural correspondence between MAP,
rTMD-F, and Morph paths in PSA was not recapitulated in NCA. On the
other hand, the ANMP paths, which were reasonably similar to iENM in PSA ($1.4\
\mathrm{\AA} \leq \delta_{F} \leq 2.7\ \mathrm{\AA}$ in Fig.~\ref{fig:m_psa})
but fairly different from Morph ($2.8\ \mathrm{\AA} \leq \delta_{F} \leq 3.1\
\mathrm{\AA}$), appeared fairly similar to both iENM and Morph in NC space.

Comparison of the NC projections of rTMD-S and rTMD-F indicated that the
  pulling velocity in rTMD directly affected the degree to which native contacts
  were broken and reformed. Consequently, different transition pathways were
  followed, as indicated by PSA, where rTMD-S clustered with the other dynamical
  methods and rTMD-F was most similar to LinInt and Morph.

NCA identified the same general division between the dynamical and non-dynamical
methods as PSA, while some subdivisions within the dynamical/non-dynamical
dichotomy are also borne out by both analyses, such as the closer grouping of
MDdMD/rTMD-S/DIMS than FRODA/DIMS or FRODA/MDdMD. However, the cusp-like
feature and overall qualitative similarity of the MENM and MAP trajectories in
NC space that set them apart from the other non-dynamical methods is apparently
not captured in PSA. The NC projection did not offer a clear hint as to why
ANMP, iENM, and MENM-SD/SP were subdivided as they were in PSA and why GOdMD
appeared as an outlier---two questions addressed by the following analysis of
the transition paths projected onto \zr and angle-angle coordinates.

\paragraph{Projections into \zr and angle-angle space.}
PSA and NCA are both general transition path analysis methods that do not
require knowledge of any system-specific order parameters or collective
variables. We employ the \zr projection (distance from and progress along the
path of linear interpolation) in order to resolve the remaining apparent
discrepancies between PSA and NCA. Because good collective variables are known
for the AdK transition~\cite{Seyler2014-uu}, we also use a 2D projection onto
domain angles~\cite{Beckstein2009-ll} to connect the conclusions derived from
the general analyses to visually intuitive structural motions of the \cto
transition.

In the \zr space projection (Fig.~\ref{fig:m_cv}C), the dynamical methods tended
to obtain the greatest distance from the LinInt reference path near the end of
the transition ($\zeta \lessapprox$~3.5~\AA) whereas the non-dynamical methods
peaked nearer the beginning. Thus, the dynamical/non-dynamical method dichotomy
previously observed in both NCA and PSA was also present in \zr space. The
structural interpretation of this behavior is, based on the projection into
angle-angle space (Fig.~\ref{fig:m_cv}B), that the dynamical methods favored a
pathway during which first the LID domains opens, followed by the NMP
domain. Non-dynamical methods produced either NMP-opening-first paths or paths
with brief LID-opening motions. In \zr space, however, dynamical methods
produced paths with a greater average and peak (orthogonal) displacement from
LinInt than non-dynamical methods (which cannot be discerned by apparent
displacements in angle-angle space), further corroborating the clusterings from
PSA.

Fast-pulling rTMD (rTMD-F), as a dynamical method, appeared as an
  exception to the dynamical/non-dynamical method dichotomy. However, both the
  projection onto domain angles and especially the \zr projection clearly showed
  that the rTMD-F path was very similar to LinInt ($\rho <$~1~\AA{} in
  Fig.~\ref{fig:m_cv}C). rTMD with very high pulling velocities of the restraint
  potential moves the system almost exlusively in the direction of the restraint
  force. For an rmsd restraint, the gradient points exactly along the LinInt
  path. Therefore, rTMD-F functions more like LinInt or Morph and less than
  equilibrium MD with an additional bias potential and hence PSA clustered rTMD-F
  with LinInt and Morph (Fig.~\ref{fig:m_psa}).

MENM-SP was the most distant member in the cluster of the four ENM-based methods
in PSA (Fig.~\ref{fig:m_psa}). Careful inspection of both angle-angle space
(Fig.~\ref{fig:m_cv}B) and \zr (Fig.~\ref{fig:m_cv}C) revealed that the MENM-SP
path contained a very large gap in the trajectory snapshots; the penultimate
conformer was located in the first half of the transition ($\zeta>$~4~\AA),
while the final snapshot was the open crystal structure end state. Such a big
gap in the path affects the discrete Hausdorff/\frechet distances because the
distance between two MENM-SP paths with well-aligned gaps is unaffected whereas
the distance between an MENM-SP path and one without gaps tends to be somewhat
larger due to large point distances originating from the latter's conformers in
the portion of the transition where the gap occurs. ANMP was also somewhat of an
outlier within the ENM cluster (Fig.~\ref{fig:m_psa}), which can be traced to
its path being much farther away from the LinInt reference than any other
ENM/Morph method ($\rho\approx 2.2$ \dots $2.3$~\AA{} versus $\rho\approx
1.3$~\AA; Fig.~\ref{fig:m_cv}C). Structurally, the NMP domain opened nearly all
the way before much of the LID motion took place, in contrast with every other
method (Fig.~\ref{fig:m_cv}B).

GOdMD produced the path with the greatest peak displacement
($\rho\approx 2.8$~\AA; Fig.~\ref{fig:m_cv}C), corresponding to complete LID
opening before substantial NMP movement occured (Fig.~\ref{fig:m_cv}B). The
results from GOdMD are unlike any of the other methods and therefore GOdMD is
well-classified as an outlier by PSA (Fig.~\ref{fig:m_psa}).

PSA was able to group fast transition path sampling methods into distinct
clusters. These groupings could be rationalized by employing projections on more
specialized collective variables. An important observation was that transition
paths were most similar to other transition paths generated by the same
method. This conclusion was, however, based on a small sample of three paths per
method. We then sought to extend our analysis to larger ensembles of paths that
would provide a statistically more meaningful comparison.

\subsection*{Comparing DIMS and FRODA transition ensembles}
We applied PSA to transition path ensembles containing hundreds of trajectories
to highlight several approaches to handling the statistical nature of dynamical
path-sampling methods and illustrate the portability of our analyses to other
systems. Ensembles of the AdK and DT \cto transitions were analyzed. DT was
selected in part to make contact with a previous study by Farrell et
al.~\cite{Farrell2010-wh} as well as provide a more challenging example to
demonstrate the ease with which PSA can filter erroneous trajectories from an
ensemble. We focused on two methods, DIMS MD and FRODA, because they differ
fundamentally in their energetic considerations yet still share several salient
features: Heavy-atom representations were used for both methods for both AdK and
DT. Both methods can generate path ensembles by employing a form of stochastic
dynamics, and they drive transitions (toward a target structure) with similar
rmsd-to-target progress variables (DIMS uses the heavy-atom rmsd-to-target for
the soft-ratcheting coordinate; FRODA attempts to gradually decrease the \Ca
rmsd to the target). Furthermore, our in-house implementations of DIMS MD
methods allowed us to efficiently generate large numbers of transitions. Four
unique ensembles and 800 total trajectories were generated: 200 pathways per
method per protein. Details about trajectory alignment for both AdK and DT is
provided in \nameref{S6_Text} of the Supporting Information.

For AdK, transition path trajectories generated with DIMS formed one cluster
that was distinct from a second cluster containing all FRODA trajectories (see
\nameref{S8_Fig} in the Supporting Information). The mean \frechet distance
$\langle\delta_{F}\rangle$ between DIMS and FRODA trajectories was $\unit{2.9\pm
  0.1}{\angstrom}$, significantly higher than the mean within the FRODA
($\unit{2.2\pm 0.1}{\angstrom}$) and DIMS ensemble ($\unit{1.4\pm
  0.2}{\angstrom}$). DIMS generated paths with smaller \frechet distances among
themselves than FRODA, while paths produced by a given method were notably more
similar among themselves than when compared with paths from the other method,
with no difference between \frechet and Hausdorff distance
(\nameref{S10_Fig} A). These observations imply that while FRODA produced paths
that sampled a larger region of AdK's configuration space than DIMS, each method
generated a unique pathway that can be viewed as a tube in configuration space
whose diameter was smaller than the typical distance between the tubes.

\begin{figure}[tb]
\begin{adjustwidth}{0.0in}{0.5in}
  \centering
  \includegraphics[scale=1]{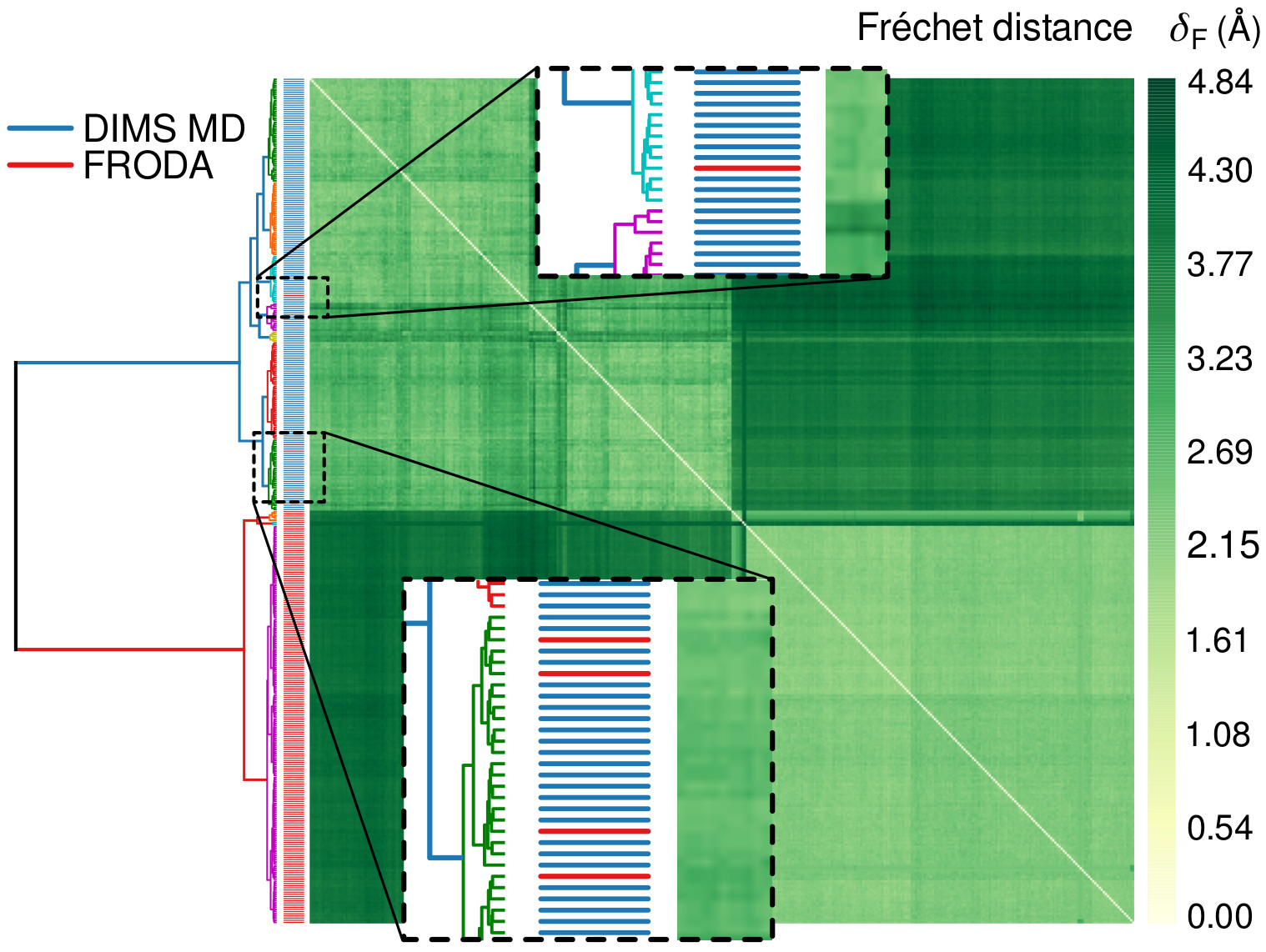}
  \caption{Clustered heat map comparing ensembles of diphtheria toxin (1MDT to
    1DDT) transition pathways produced by DIMS (blue bars) and FRODA (red bars)
    using the \frechet distance \dF. All clusterings are produced using
    the Ward's criterion in ascending distance order; incomplete trajectories
    were filtered and not displayed (see text). The insets show that five FRODA
    trajectories cluster together with the DIMS ensemble.
    }
  \label{fig:e_psa}
\end{adjustwidth}
\end{figure}

While the AdK analysis was relatively straightforward, the DT heat map
immediately revealed nine erroneous FRODA trajectories producing \frechet
distances upwards of 5~\AA{} from any other path (see \nameref{S9_Fig} for the
original clustering). Erroneous paths were removed by specifying a distance
cutoff and re-clustering using the trimmed FRODA ensemble. Visual inspection of
the omitted trajectories confirmed that they either stopped short of the target
or that they came somewhat near the target but continued to dramatically wander
in its vicinity. All the DIMS trajectories and most of the FRODA trajectories
formed two large, separate clusters (Fig.~\ref{fig:e_psa}). Interestingly, five
FRODA paths were among the cluster of DIMS paths (Fig.~\ref{fig:e_psa} insets)
whereas there was no intermixing between DIMS and FRODA among AdK trajectories
(\nameref{S8_Fig}). Furthermore, the DIMS--DIMS, FRODA--FRODA, and DIMS--FRODA
distributions of \frechet (and Hausdorff) distances for the DT
  transitions overlap (\nameref{S10_Fig} B), which indicates (via the triangle
inequality of the metric) that DIMS and FRODA paths are situated in each others'
vicinity.

The observation of intermixing supports the idea that the geometrical targeting
procedure of FRODA is able to sample the space of trajectories accessible to the
force field-based DIMS MD. As FRODA is guaranteed to produce stereochemically
correct transitions without regards to the energetics, the set of possible paths
that it can produce ought to be a superset of all trajectories produced by other
sampling algorithms that, in addition to preserving stereochemistry, implement
more detailed energetics that would further constrain the sampling space. The
finding that several FRODA trajectories cluster with the DT DIMS ensemble hints
at the possibility that there exists a hierarchy of accessible path spaces that
can be sampled by different methods depending on the energetics. 

The fact that no intermixing was observed in the AdK transitions does not
  necessarily contradict the hypothesis that FRODA trajectory space contains
  DIMS trajectory space, because the specifics of the AdK transition and the
  nature of the FRODA sampling algorithm seem to make it less likely that FRODA
  would actually sample a DIMS pathway (as discussed in more detail in
  \nameref{S7_Text}). It is not immediately obvious how one would tune one
  particular path-generating algorithm (such as FRODA) in order to increase its
  likelihood to produce a path characteristic of another algorithm (such as
  DIMS), although we already observed that the variation of the pulling speed in
  the rTMD method led to quantitatively (Fig.~\ref{fig:m_psa}) and qualitatively
  (Fig.~\ref{fig:m_cv}) different paths. In particular, fast pulling (rTMD-F)
  generated paths similar to linear interpolation (LinInt), whereas slow pulling
  (rTMD-S) paths were more similar to DIMS and MDdMD. Thus, although we do not
  yet in general understand the relationship between the pathways sampled by
  different algorithms, PSA appears to be a useful tool to tackle this
  question.

The ultimate goal is, of course, to find a method that reliably samples
transitions realized in the real system. The analysis presented here
should also aid in identifying the overlap between different
sampling methods and experimental data (e.g.\ from femtosecond structural
biology experiments) when such data become available.

\subsection*{Hausdorff pairs connect PSA to molecular detail}

PSA is a general approach that can operate on the full \tnd trajectories without
requiring any system-specific knowledge. It provides a very broad means to
categorize transitions as distinct from one other. But as described so far, it
is difficult to relate the global PSA analysis to physically relevant
differences at the molecular level. To address this question we introduce the
new concept of ``Hausdorff pairs'' (or ``\frechet pairs'') that allows us to
pinpoint conformations that may be more likely to exhibit geometric (structural)
features relevant to conformational change.

By construction, the Hausdorff and the \frechet distances identify a point-wise
distance between two particular conformers, one on each path, as the global
distance between the paths. The path metrics therefore induce a map between
a conformer on one path to a conformer on another whose separation distance
is, in some sense, a maximal deviation between the paths. We term such a pair
of conformers a \emph{Hausdorff pair} (\dH-pair) or a \emph{\frechet pair}
(\dF-pair). These conformers can be examined at the molecular or atomic level
to reveal the specific structural discrepancies that give rise to large
deviations in configuration space between pairs of paths.

\begin{figure}[]
\begin{adjustwidth}{-2.0in}{0.5in}
  \centering
  \includegraphics[]{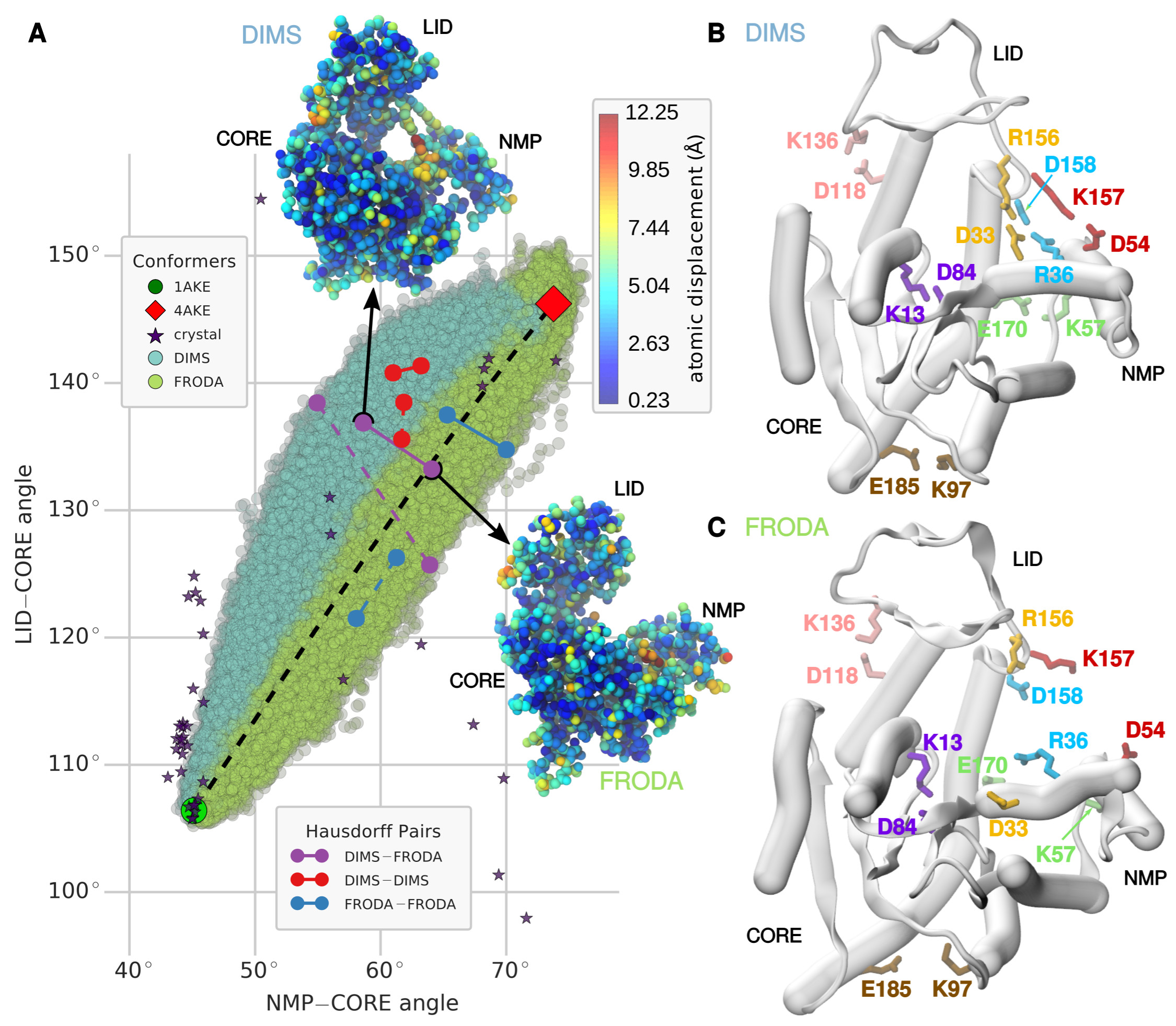}
  \caption{``Hausdorff pairs'' ($\delta_H$-pairs) analysis using 200 DIMS (cyan)
    and 200 FRODA (light green) trajectories projected into AA space.  Hausdorff
    distances were computed for all unique path pairs.  (A) Conformer
    pairs---corresponding to the $\delta_H$-pairs with the median and maximum
    Hausdorff distances (solid and dashed lines, respectively)---are projected
    onto the domain angle space for the following comparisons: DIMS--FRODA
    (purple), DIMS--DIMS (red), and FRODA--FRODA (blue). Experimental
      crystal structures, including some intermediates, are shown as stars
      \cite{Beckstein2009-ll}, with further details  available in
      \nameref{S1_Tab}. Insets: Two heavy-atom representations are shown for
    the median $\delta_H$-pair between a DIMS path and FRODA path, corresponding
    to snapshots from the respective trajectories. The magnitude of the
    displacement vector between the two conformations is projected onto each
    atom. Color bar units for atomic displacement are in Ångström. The initial
    and final conformations (green circle and red diamond, respectively) are
    shown along with the linear interpolation path LinInt -- black dashed line)
    for reference.  (B,C) Salt bridges in the DIMS and FRODA conformers from the
    DIMS-FRODA median Hausdorff pair. Three LID-NMP salt bridges (R156-D33,
    D158-R36, and K157-D54) and a CORE-NMP salt bridge (E170-K57) are intact in
    the DIMS structure (B) that are broken in the FRODA structure (C).  The
    residues responsible for these salt bridges tug on the NMP domain more
    substantially than their counterparts in the LID domain, which are located
    toward the base of the LID.}
  \label{fig:dhp}
\end{adjustwidth}
\end{figure}

As an explicit example, we identified three Hausdorff pairs for the DIMS and
FRODA \cto AdK transition ensembles and projected them in AA space
(Fig.~\ref{fig:dhp}A). We first segregated the full set of Hausdorff distance
measurements into: (1) mutual distances among DIMS paths, (2) mutual distances
among FRODA paths, and (3) inter-method distances measured between a DIMS and
a FRODA path. A total of $N(N-1)/2=79800$ \dH-pairs were identified for the
ensemble of $N=400$ paths. In order to present representative data for the
whole ensemble, we identified the two \dH-pairs associated with the median and
maximum Hausdorff distances for each comparison (1), (2), and (3) as defined
above.

As a typical example we explicitly examined the median \dH-pair identified for
the inter-method comparisons and projected the atomic displacements onto each
structure to locate regions of large deviation (see structures in
Fig.~\ref{fig:dhp}A). It became apparent that the NMP domain in the DIMS
structure was closer to the LID domain because a number of evolutionary
conserved salt bridges (D33--R156, R36--D158, D54--K157) persisted late into the
transition due to the strong electrostatic interaction between the acidic and
basic moieties~\cite{Beckstein2009-ll} (Fig.~\ref{fig:dhp}B). FRODA, on the
other hand, operating on purely geometric principles and neglecting Coulomb
interactions, does not account for the influence of salt bridges on the
transition and the associated \dH-pair structure exhibited broken salt bridges
in the inter-LID/NMP region (Fig.~\ref{fig:dhp}C). It is therefore not
surprising that the FRODA trajectory did not show the ``salt-bridge
zipper''~\cite{Beckstein2009-ll}, which manifested itself as discerning
difference between the DIMS and FRODA trajectories. With salt bridges located
across the NMP domain but primarily on the side of the LID, the LID is
relatively free to move to an open configuration, whereas the NMP domain is
prevented from fully opening until the salt bridges are broken. These
considerations are consistent with the tendency of DIMS paths to primarily favor
a LID-opening pathway (Fig.~\ref{fig:dhp}A, blue circles), while FRODA paths
(Fig.~\ref{fig:dhp}A, green circles) sampled the region around LinInt
(Fig.~\ref{fig:dhp}A, black dashed line) corresponding to simultaneous
LID/NMP-opening.

A Hausdorff pair describes the two frames at which the two trajectories
  in question differ most. Additionally, the regions where trajectories differ
  to varying degrees from each other might also be of interest. This kind of
  information is provided by the the set of nearest neighbor distances along a
  path. Eq.~\ref{eq:direct-dh} defines the \emph{nearest neighbor distance} of
  point $p_{k}$ on path $P$ from path $Q$ as $\delta_{h}(k; P\mid Q) :=
  \delta_{h}(p_{k} \mid Q) := \min_{q\in Q} d(p_{k}, q)$ and the nearest
  neighbor distance of point $q_{k}$ on path $Q$ from path $P$ as $\delta_{h}(k;
  Q \mid P)$. In general, these two distances are not symmetric, i.e.{}
  $\delta_{h}(k; P \mid Q) \neq \delta_{h}(j; Q \mid P)$ for any conformations
  $j, k$. When $\delta_{h}\big(k(\xi); P \mid Q\big)$ and
  $\delta_{h}\big(j(\xi); Q \mid P\big)$ are plotted against a suitable common
  order parameter $\xi$, the regions of large and small differences between
  trajectories can be quantified. For example, in \nameref{S11_Fig}, the nearest
  neighbor distances of the three pairs of trajectories corresponding to the
  median Hausdorff pairs in Fig.~\ref{fig:dhp}A showed that the DIMS and FRODA
  trajectories primarily differed in the first $\sim 60$\% of the transition,
  which corresponds to LID-opening in DIMS and simultaneous LID/NMP-opening in
  FRODA. The DIMS trajectories differed almost uniformly along the whole path by
  only $\lessapprox 1.3$~\AA, suggesting that they follow a similar path
  perturbed by thermal fluctuations. The FRODA trajectories differed by $\sim
  2$~\AA{} during the middle half of the transition but practically coincided at
  beginning and end, showing that FRODA can accurately connect two given
  endpoint structures even with its stochastic component enabled.

The Hausdorff-pair and nearest neighbor distance analysis naturally
followed from the formulation of PSA. Even though only \Ca atoms were used to
distinguish DIMS from FRODA trajectories and hence the level of detail of PSA
was primarily restricted to conformational differences in the protein backbone,
atomic-scale analysis of Hausdorff-pairs was able to reveal the molecular
determinants responsible for the structural differences.

\subsection*{Conclusions}

\paragraph{Summary.} We developed a flexible and
quantitative framework for analyzing macromolecular
transition paths using path metrics as a means to measure the mutual similarity
of paths in configuration space, potentially using the full $3N$-dimensional
configuration space information. As far as we are aware, there is currently no
standard procedure for quantitatively analyzing and characterizing transition
paths. After comparing a set of transitions from a variety of
path-sampling algorithms and analyzing transition ensembles of two sets of
dynamical, stochastic trajectories, PSA's viability as
a tool to quantitatively compare transition paths
appears promising.

In particular, PSA demonstrated that for the AdK \cto transition, fast path
sampling methods generated paths that were more similar to other paths produced
by the same method than to any other paths, suggesting that among these
methods there is currently no real consensus for what a realistic conformational
transition looks like. Hierarchical clustering in combination with a heat map
representation indicated broad patterns whereby dynamical methods tended to be
clustered with each other while non-dynamical ones (especially most of the
ENM-based ones) were more similar to each other than to methods such as DIMS or
FRODA. The clustering was qualitatively confirmed by a range of low-dimensional
projections on collective variables, which can be used synergistically with PSA
once additional knowledge about the system of interest is available.

Analysis of ensembles of \cto transition of DT produced with the computationally
very distinct DIMS and FRODA methods showed that in about 2.5\% of the cases,
FRODA sampled a similar region of path space as DIMS. The finding suggests a
picture of path space where different methods preferentially generate
trajectories in their own region even though in rare cases neighboring regions
are sampled. It will be of interest to determine regions of this space where
multiple methods overlap and evaluate such consensus regions as candidates for
more realistic transition pathways. The ensemble analysis also clearly showed
that at least for large scale macromolecular transitions, the \frechet and the
Hausdorff distance are equally appropriate measures for path similarity, with
the Hausdorff distance being cheaper to compute.

A key advantage of PSA is its generality in that no system specific knowledge is
required and all trajectory data can easily be used. That generality might,
however, obscure the physical and biological detail that is important for a
mechanistic understanding of protein function. The new concept of Hausdorff (or
\frechet) pairs within the context of PSA suggests an effective
approach to identifying the molecular determinants responsible for differences in paths
between sampling methods, which could be related to variations in
protein function, differences in path-sampling algorithms/models, or some
combination of both (as demonstrated for the comparison between DIMS and FRODA).

\paragraph{Future directions.} The rmsd proved a useful choice for the point
metric to measure structural similarity as its preponderance in the literature
helps to connect path metric distance measurements to familiar intuitions
although its interpretation is not entirely obvious across disparate
contexts~\cite{Maiorov1994-od}. A further study would, for instance, benefit
from a revised definition of native contacts where consecutive alpha carbons
(within the cutoff distance) would be ignored. As such, a revised native- or
self-contacts measure could be used as a point metric instead of rmsd. We also
plan to employ k-medoids clustering---used to identify a ``median'' element
(i.e., a medoid) in a set---as one possible approach to identifying
representative transition paths, bringing us a step closer to identifying
transition tubes or candidates for reaction coordinates.

Furthermore, PSA can aid the assessment of enhanced path-sampling algorithms and
their performance by quantifying the degree of similarity to gold-standard
transition paths (for instance, equilibrium MD transitions or---when
available---experimental time-resolved structural data). Such quantitative
comparisons would be key to assess the physical plausibility of transitions
generated by enhanced sampling methods. However, comparison of such
  transition paths to true equilibrium paths will require the development of a
  method to identify the actual transition events in the equilibrium data and to
  distinguish them from residency in (meta) stable states, ideally without
  introducing problem-specific order parameters. Possible approaches may include
  an analysis of the distribution of Hausdorff nearest neighbor distances or
  discrete \frechet couplings or techniques to match subtrajectories
  \cite{Buchin:2011zr, Sankararaman:2013ys} but identifying barrier crossing
  events from coordinate trajectories in a general manner alone remains an open
  problem. By
extending the analysis of Hausdorff pairs we should also be able to better
pinpoint key structural events or mutations that affect the function of
biomolecules and so improve our understanding of the connection between
  protein structure, dynamics and function.
  
\section*{Supporting Information}


\subsection*{S1 Text}
\label{S1_Text}
\textbf{Alternative distance functions: average Hausdorff and discrete average \frechet.}
This text explores variations of the Hausdorff and discrete \frechet metrics
based on averages rather than maxima, which are intended to reduce sensitivity
to outliers. Definitions are provided and are shown, via example, to violate
the triangle inequality. We repeat the path-sampling methods comparison and
discuss their behavior in the context of PSA.

\subsection*{S2 Text}
\label{S2_Text}
\textbf{Comments on the numerical implementation of path metrics.}
We informally discuss and comment on the numerical aspects of computing
transition path similarity and the algorithms used to calculate Hausdorff and
\frechet distances.

\subsection*{S3 Text}
\label{S3_Text}
\textbf{Comments on the selection and validation of clustering algorithms.}
In this text we mention qualitative and quantitative
considerations in selecting the Ward linkage criterion for
hierarchical clustering, in the context of other linkage criteria.
Some comments on potential data interpretation pitfalls when
performing general cluster analyses are provided with a view
toward viable approaches to PSA cluster and data validation.

\subsection*{S4 Text}
\label{S4_Text}
\textbf{Mathematical details for the energetics and dynamics of the double-barrel model.}
Mathematical details are provided for the simulation
of the double-barrel model. The system assumes overdamped Langevin dynamics (Brownian
motion) and numerical integration was performed using a first-order scheme in time.
The construction of the model permits consistent coarse-graining with respect to the
number of particles in a cluster, effectively allowing tuning of the number of
degrees of freedom, or the dimensionality of the configuration space.

\subsection*{S5 Text}
\label{S5_Text}
\textbf{Expanded overview of the transition path generating algorithms
used in this study.}
Here we provide a short review---for reader convenience---of the transition path generating
algorithms used in the comparison of sampling methods. We summarize the key
aspects of each of the physical models and path generating algorithms to help
lay the groundwork for connecting algorithmic/model differences
to differences between the respective transition paths that were produced.

\subsection*{S6 Text}
\label{S6_Text}
\textbf{Details of structural alignment procedures for protein alignment
prior to path similarity analysis.}
This text summarizes the considerations involved in structurally
aligning conformer snapshots prior to running path similarity analysis on a
set of transition paths. We provide specific details and motivations as to
the alignment procedures used for AdK and DT trajectories.

\subsection*{S7 Text}
\label{S7_Text}
\textbf{Discussion of FRODA trajectories as a superset of other
  trajectory-generating algorithms.} %
In this text we qualitatively discuss the difference between the adenylate
kinase (AdK) and diphtheria toxin transition (DT). These differences rationalize
the observation that for the DT transition some FRODA trajectories overlap with
the ensemble of DIMS trajectories whereas no such overlap was observed in the
case of AdK.

\subsection*{S1 Fig}
\label{S1_Fig}
\textbf{Effect of temperature and dimensionality on the distribution of path
  metrics.}
Violin plots \cite{Hintze:1998tw} show the distributions of discrete \frechet
distances for double-barrel simulations of one particle (orange) and eight
particles (purple) for temperatures ranging between 0 and \unit{500}{\kelvin} in
\unit{50}{\kelvin} increments (panels A--K) and at
\unit{600}{\kelvin} (panel L). Black points correspond to individual
\frechet distance measurements, with distance units in nm rmsd. A kernel density
estimate (kde) is shown for each $N$, $T$ pair to qualitatively emphasize the
behaviors of the distributions across the entire temperature range; the
bandwidth for each pair is explicitly set to produce two distinct distributions
at low temperatures and gradually increased to generate smooth, single
distributions at high temperatures. The separated distributions at low
temperatures merge between \unit{300}{\kelvin} to \unit{450}{\kelvin}, with the
eight-particle simulations merging toward higher temperatures relative to the
one-particle simulations.

\subsection*{S2 Fig}
\label{S2_Fig}
\textbf{Correlation analysis of \frechet and Hausdorff distances in the toy model.}
Regression analyses examining the correlation between corresponding \frechet
(horizontal axes) and Hausdorff (vertical axes) distance measurements are
plotted along with the joint distributions plots for double-barrel simulations
of one particle (orange points) and eight particles (purple) for temperatures
ranging between 0 and \unit{500}{\kelvin} in \unit{50}{\kelvin} increments
(panels A--K) and at \unit{600}{\kelvin} (panel L). Scatter points correspond
to individual \frechet distance
measurements in nm rmsd and are plotted with the line produced by linear
regression.  The shading about the regression lines correspond to a 95\%
confidence interval.  Kernel density estimates (kde) are shown for each $N$, $T$
pair and are computed using the same set of bandwidth constants specified in
\nameref{S1_Fig}. The separated distributions at low temperatures merge between
\unit{300}{\kelvin} to \unit{450}{\kelvin}, with a notable narrowing of the
range of distance measurements occuring between \unit{400}{\kelvin} to
\unit{450}{\kelvin}.

\subsection*{S3 Fig}
\label{S3_Fig}
\textbf{Effect of temperature and dimensionality on the correlation between
  \frechet and Hausdorff distance.}
Coefficients of the Pearson correlation between Hausdorff and \frechet distances
for one- and eight-particle simulations plotted as a function of
temperature. Path distances remain well correlated up to \unit{300}{\kelvin} and
are least correlated at \unit{500}{\kelvin}, with the one-particle simulations
exhibiting a substantially larger drop in correlation. At the highest
temperature the central barrier becomes negligible and the simulations start to
equally sample a single tube dominated by the steep repulsive walls. Therefore,
the paths become more similar again between the $N=1$ and $N=8$ clusters and the
correlation coefficient increases.

\subsection*{S4 Fig}
\label{S4_Fig}
\textbf{Temperature-dependent transition from two to one distinct paths in the
  toy model.}
The means and standard deviations of the discrete \frechet (blue) and Hausdorff
(red) distances for double-barrel simulations of one particle (A) and
eight particles (B) are shown as functions of temperature.
Measurements for simulations at \unit{250}{\kelvin} and below were divided into
an upper and lower distribution by separating distance measurements above and
below a \unit{1.25}{\nano\meter} cutoff. Above the temperature cutoff, all
measurements were treated as part of the same distribution. Both the \frechet
and Hausdorff metric lose the ability to distinguish between the two barrels as
the paths begin to wander out of well-defined pathways when the temperature is
on the order of the equivalent energy of the central barrier ($2 {k_{B}T}$ at
\unit{300}{\kelvin}). At higher temperatures, thermal perturbations become large
relative to the barrier, permitting particle clusters to explore the full width
of the potential spanning both barrels so as to generate trajectories confined
to a single, unified pathway.

\subsection*{S5 Fig}
\label{S5_Fig}
\textbf{Annotated \frechet distance matrix of AdK transition trajectories
  generated by different path-sampling methods.}
The \frechet distance matrix from Fig.~\ref{fig:m_psa} is shown with the
numerical values of $\delta_{F}$ (rounded to one decimal) superimposed. Due to
the size of the distance matrix, the high resolution image is provided as a
simple means for online data exploration with the help of the zoom function of
an image viewer.

\subsection*{S6 Fig}
\label{S6_Fig}
\textbf{Influence of the linkage algorithm on the clustered PSA comparison of
  different path sampling methods.}
Different linkage algorithms were used to cluster the \frechet distances
produced by path-sampling methods for the AdK \cto transition. Smaller distances
(in units of \AA{}~rmsd) indicate transition paths with greater
similarity. Dendrograms for each heat map correspond to the hierarchical
clustering produced by the single (A), complete (B), average
(C), and weighted (D) linkage algorithms, and depict a
hierarchy of clusters with smaller node heights of parent clusters indicating
greater similarity between child clusters.

\subsection*{S7 Fig}
\label{S7_Fig}
\textbf{PSA comparison of different path sampling methods based on the Hausdorff
  distance.}
(A) Heat map for path-sampling methods for the AdK \cto transition of
Hausdorff distances produced using the Ward algorithm. Clusters are identical
to the Ward clustering for \frechet distances in Fig.~\ref{fig:m_psa}.
(B) Correlation and joint distributions between discrete \frechet versus
Hausdorff distance measurements (in \AA{}~rmsd) for the AdK \cto methods
comparison. Strong linear correlation indicated
by the scatter plot, with a Pearson correlation coefficient very close to
unity, indicates that either metric could have been used to perform the
path-sampling methods analysis with essentially identical results. A
slight deviation of the scatter points below the line of unity slope
is consistent with the fact that \frechet distances are bounded from
below by corresponding Hausdorff distances.

\subsection*{S8 Fig}
\label{S8_Fig}
\textbf{Clustered PSA heat map of AdK transition ensembles.}
Clustered heat map comparing path ensembles of adenylate kinase (1AKE:A to
4AKE:A) transition paths produced by DIMS (red bars) and FRODA (blue bars)
using the discrete \frechet distance \dF. Clustering was produced using
the Ward algorithm in ascending distance order.

\subsection*{S9 Fig}
\label{S9_Fig}
\textbf{Clustered PSA heat map of raw DT transition ensembles.}
Clustered heat map comparing the raw path ensembles of diphtheria toxin
(1MDT:A to 1DDT:A) transition paths produced by DIMS (red bars) and FRODA
(blue bars) using the discrete \frechet distance \dF. Clustering was produced
using the Ward algorithm in ascending distance order. Nine erroneous FRODA paths
(orange cluster) were very distant from all other paths---all nine were
removed from the ensemble and to produce the heat map dendrogram in
Fig.~\ref{fig:e_psa}.

\subsection*{S10 Fig}
\label{S10_Fig}
\textbf{Correlation between \frechet and Hausdorff distance in ensemble
  comparisons.}
Correlations and joint distributions of discrete \frechet versus Hausdorff
distance measurements (in \AA{}~rmsd) of the AdK (A) and DT
(B) ensemble analyses are shown. Measurements are divided into three
separate distributions: (1) mutual distances among DIMS paths (red), (2) mutual
distances among FRODA paths (green), and (3) inter-method distances measured
between a DIMS and a FRODA path (blue). Both scatter plots show strong
correlation between the path metrics for all the distributions, with Pearson
correlation coefficients equal to unity and p-values equal to zero to two
decimal places, indicating that either metric could have been used to perform
the path-sampling methods analysis to obtain essentially identical results. A
slight deviation of the scatter points below the line of unity slope is
consistent with the fact that \frechet distances are bounded from below by
corresponding Hausdorff distances. DIMS simulations exhibited less variation
than FRODA in the AdK transition, but had a larger average variation in the DT
transition. In both cases, inter-method DIMS-FRODA comparisons were
substantially larger than comparisons of pairs of paths from a single method.

\subsection*{S11 Fig}
\label{S11_Fig}
\textbf{Nearest neighbor distances along trajectories for the median Hausdorff
  pairs in the AdK ensemble comparison.}  The nearest neighbor distances
$\delta_{h}(k; Q \mid P)$ (solid line, \solidrule) and $\delta_{h}(k; P \mid Q)$
(dashed line, \dashedrule) between pairs of paths $P$/$Q$ belonging to the three
median Hausdorff pairs in the AdK ensemble comparison (Fig.~\ref{fig:dhp}A) are
shown for DIMS/FRODA (purple), DIMS/DIMS (blue), and FRODA/FRODA (green). The
largest value $\max_{k,j}\big(\delta_{h}(k; Q \mid P), \delta_{h}(j; P \mid Q)\big)$ is
the actual Hausdorff distance. For illustration purposes, nearest neighbor
distances are plotted as a function of frame number $k$ normalized to the
interval $[0,1]$ (i.e., $k/|P|$), where $0$ ($1$) corresponds to the first
(last) frame. In general, an appropriate one-dimensional order parameter should
be chosen in order to plot nearest neighbor distances for structurally
corresponding trajectory frames.

\subsection*{S1 Table}
\label{S1_Tab}
\textbf{Numerical values for domain angles for experimental crystal structures
  of AdK.} The NMP-CORE angle and LID-CORE angle for \textit{E. coli} EcoAdK or
models of EcoAdK based on crystal structures of homologous proteins
\cite{Beckstein2009-ll} was computed from the \Ca atoms in the NMP, LID
and CORE domain as described in Methods. The angles are plotted in
Fig.~\ref{fig:dhp}A.

\section*{Acknowledgments}
We thank David Dotson for helpful discussions and Marc Delarue for
introducing us to native contact analysis for conformational
transitions. This work used the Extreme Science and Engineering Discovery
Environment (XSEDE), which is supported by National Science Foundation grant
number ACI-1053575 (allocation MCB130177 to OB). SLS was supported in part by
a Wally Stoelzel Fellowship from the Department of Physics at Arizona State
University. AK was funded in part by the ARCS Foundation.

%
%
%
%
%

\bibliography{paperpile,manually_added,revision}

\begin{thebibliography}{100}

\bibitem{Yon1998-jp}
Yon JM, Perahia D, Gh\'{e}lis C.
\newblock Conformational dynamics and enzyme activity.
\newblock Biochimie. 1998 Jan;80(1):33--42.

\bibitem{Karplus2005-vz}
Karplus M, Gao YQ, Ma J, van~der Vaart A, Yang W.
\newblock Protein structural transitions and their functional role.
\newblock Philos Trans A Math Phys Eng Sci. 2005 15~Feb;363(1827):331--55;
  discussion 355--6.

\bibitem{Henzler-Wildman2007-bp}
Henzler-Wildman K, Kern D.
\newblock Dynamic personalities of proteins.
\newblock Nature. 2007 13~Dec;450(7172):964--972.

\bibitem{Dror:2012cr}
Dror RO, Dirks RM, Grossman JP, Xu H, Shaw DE.
\newblock Biomolecular simulation: a computational microscope for molecular
  biology.
\newblock Annu Rev Biophys. 2012;41:429--52.

\bibitem{Orozco:2014dq}
Orozco M.
\newblock A theoretical view of protein dynamics.
\newblock Chem Soc Rev. 2014;43:5051--5066.

\bibitem{Schwartz2009-xm}
Schwartz SD, Schramm VL.
\newblock Enzymatic transition states and dynamic motion in barrier crossing.
\newblock Nat Chem Biol. 2009 Aug;5(8):551--558.

\bibitem{Lei2007-hq}
Lei H, Duan Y.
\newblock Improved sampling methods for molecular simulation.
\newblock Curr Opin Struct Biol. 2007 Apr;17(2):187--191.

\bibitem{Yang2008-bi}
Yang LW, Chng CP.
\newblock Coarse-grained models reveal functional dynamics--I. Elastic network
  models--theories, comparisons and perspectives.
\newblock Bioinform Biol Insights. 2008 4~Mar;2:25--45.

\bibitem{Chng2008-wc}
Chng CP, Yang LW.
\newblock Coarse-grained models reveal functional {dynamics--II}. Molecular
  dynamics simulation at the coarse-grained level--theories and biological
  applications.
\newblock Bioinform Biol Insights. 2008 5~Mar;2:171--185.

\bibitem{Zuckerman2011-ts}
Zuckerman DM.
\newblock Equilibrium sampling in biomolecular simulations.
\newblock Annu Rev Biophys. 2011;40:41--62.

\bibitem{Christen2008-if}
Christen M, van Gunsteren WF.
\newblock On searching in, sampling of, and dynamically moving through
  conformational space of biomolecular systems: A review.
\newblock J Comput Chem. 2008 30~Jan;29(2):157--166.

\bibitem{Seyler2014-uu}
Seyler SL, Beckstein O.
\newblock Sampling large conformational transitions: adenylate kinase as a
  testing ground.
\newblock Mol Simul. 2014 9~Aug;40(10-11):855--877.

\bibitem{Schlitter1994-xz}
Schlitter J, Engels M, Kr{\"{u}}ger P.
\newblock Targeted molecular dynamics: a new approach for searching pathways of
  conformational transitions.
\newblock J Mol Graph. 1994 1~Jun;12(2):84--89.

\bibitem{Voter1997-lc}
Voter AF.
\newblock Hyperdynamics: Accelerated Molecular Dynamics of Infrequent Events.
\newblock Phys Rev Lett. 1997 19~May;78(20):3908--3911.

\bibitem{Woolf1998-xh}
Woolf TB.
\newblock Path corrected functionals of stochastic trajectories: towards
  relative free energy and reaction coordinate calculations.
\newblock Chem Phys Lett. 1998 19~Jun;289(5--6):433--441.

\bibitem{Sugita1999-de}
Sugita Y, Okamoto Y.
\newblock Replica-exchange molecular dynamics method for protein folding.
\newblock Chem Phys Lett. 1999 26~Nov;314(1--2):141--151.

\bibitem{Laio2002-qx}
Laio A, Parrinello M.
\newblock Escaping free-energy minima.
\newblock Proc Natl Acad Sci U S A. 2002 1~Oct;99(20):12562--12566.

\bibitem{Hamelberg2004-ot}
Hamelberg D, Mongan J, McCammon JA.
\newblock Accelerated molecular dynamics: a promising and efficient simulation
  method for biomolecules.
\newblock J Chem Phys. 2004 22~Jun;120(24):11919--11929.

\bibitem{Kubitzki2008-qy}
Kubitzki MB, de~Groot BL.
\newblock The atomistic mechanism of conformational transition in adenylate
  kinase: a {TEE-REX} molecular dynamics study.
\newblock Structure. 2008 6~Aug;16(8):1175--1182.

\bibitem{Barnett2009-xl}
Barnett CB, Naidoo KJ.
\newblock Free Energies from Adaptive Reaction Coordinate Forces ({FEARCF)}: an
  application to ring puckering.
\newblock Mol Phys. 2009 20~Apr;107(8-12):1243--1250.

\bibitem{Abrams2013-hh}
Abrams C, Bussi G.
\newblock Enhanced Sampling in Molecular Dynamics Using Metadynamics,
  {Replica-Exchange}, and {Temperature-Acceleration}.
\newblock Entropy. 2013 27~Dec;16(1):163--199.

\bibitem{Bolhuis2002-hj}
Bolhuis PG, Chandler D, Dellago C, Geissler PL.
\newblock Transition path sampling: {Throwing} ropes over rough mountain
  passes, in the dark.
\newblock Annu Rev Phys Chem. 2002;53(1):291--318.

\bibitem{E2005-vp}
E W, Ren W, Vanden-Eijnden E.
\newblock Finite temperature string method for the study of rare events.
\newblock J Phys Chem B. 2005 14~Apr;109(14):6688--6693.

\bibitem{Maragliano2006-dy}
Maragliano L, Fischer A, Vanden-Eijnden E, Ciccotti G.
\newblock String method in collective variables: minimum free energy paths and
  isocommittor surfaces.
\newblock J Chem Phys. 2006 14~Jul;125(2):24106.

\bibitem{Van_der_Vaart2007-zp}
van~der Vaart A, Karplus M.
\newblock Minimum free energy pathways and free energy profiles for
  conformational transitions based on atomistic molecular dynamics simulations.
\newblock J Chem Phys. 2007 28~Apr;126(16):164106.

\bibitem{Pan2008-eg}
Pan AC, Sezer D, Roux B.
\newblock Finding transition pathways using the string method with swarms of
  trajectories.
\newblock J Phys Chem B. 2008 20~Mar;112(11):3432--3440.

\bibitem{Jonsson1998-dz}
J\'{o}nsson H, Mills G, Jacobsen KW.
\newblock Nudged elastic band method for finding minimum energy paths of
  transitions.
\newblock In: Berne BJ, Ciccoti G, Coker DF, editors. Classical and Quantum
  Dynamics in Condensed Phase Simulations. World Scientific; 1998. p. 385--394.

\bibitem{Henkelman2000-fm}
Henkelman G, J\'{o}nsson H.
\newblock Improved tangent estimate in the nudged elastic band method for
  finding minimum energy paths and saddle points.
\newblock J Chem Phys. 2000 8~Dec;113(22):9978--9985.

\bibitem{Henkelman2000-jy}
Henkelman G, Uberuaga BP, J\'{o}nsson H.
\newblock A climbing image nudged elastic band method for finding saddle points
  and minimum energy paths.
\newblock J Chem Phys. 2000 8~Dec;113(22):9901--9904.

\bibitem{Fischer1992-tp}
Fischer S, Karplus M.
\newblock Conjugate peak refinement: an algorithm for finding reaction paths
  and accurate transition states in systems with many degrees of freedom.
\newblock Chem Phys Lett. 1992 Jun;194(3):252--261.

\bibitem{Franklin2007-mz}
Franklin J, Koehl P, Doniach S, Delarue M.
\newblock {MinActionPath}: maximum likelihood trajectory for large-scale
  structural transitions in a coarse-grained locally harmonic energy landscape.
\newblock Nucleic Acids Res. 2007 Jul;35(Web Server issue):W477--82.

\bibitem{Tirion1996-qg}
Tirion MM.
\newblock Large Amplitude Elastic Motions in Proteins from a
  {Single-Parameter}, Atomic Analysis.
\newblock Phys Rev Lett. 1996 26~Aug;77(9):1905--1908.

\bibitem{Bahar1997-nb}
Bahar I, Atilgan AR, Erman B.
\newblock Direct evaluation of thermal fluctuations in proteins using a
  single-parameter harmonic potential.
\newblock Fold Des. 1997;2(3):173--181.

\bibitem{Atilgan2001-qf}
Atilgan AR, Durell SR, Jernigan RL, Demirel MC, Keskin O, Bahar I.
\newblock Anisotropy of fluctuation dynamics of proteins with an elastic
  network model.
\newblock Biophys J. 2001 Jan;80(1):505--515.

\bibitem{Maragakis2005-me}
Maragakis P, Karplus M.
\newblock Large amplitude conformational change in proteins explored with a
  plastic network model: adenylate kinase.
\newblock J Mol Biol. 2005 30~Sep;352(4):807--822.

\bibitem{Cortes2005-hw}
Cort\'{e}s J, Sim\'{e}on T, Ruiz~de Angulo V, Guieysse D, Remaud-Sim\'{e}on M,
  Tran V.
\newblock A path planning approach for computing large-amplitude motions of
  flexible molecules.
\newblock Bioinformatics. 2005 Jun;21 Suppl 1:i116--25.

\bibitem{Seeliger2007-ka}
Seeliger D, Haas J, de~Groot BL.
\newblock Geometry-based sampling of conformational transitions in proteins.
\newblock Structure. 2007 Nov;15(11):1482--1492.

\bibitem{Raveh2009-wa}
Raveh B, Enosh A, Schueler-Furman O, Halperin D.
\newblock Rapid sampling of molecular motions with prior information
  constraints.
\newblock PLoS Comput Biol. 2009 Feb;5(2):e1000295.

\bibitem{Farrell2010-wh}
Farrell DW, Speranskiy K, Thorpe MF.
\newblock Generating stereochemically acceptable protein pathways.
\newblock Proteins. 2010 1~Nov;78(14):2908--2921.

\bibitem{Best2013-na}
Best RB, Hummer G, Eaton WA.
\newblock Native contacts determine protein folding mechanisms in atomistic
  simulations.
\newblock Proc Natl Acad Sci U S A. 2013 29~Oct;110(44):17874--17879.

\bibitem{Balsera1996-yo}
Balsera MA, Wriggers W, Oono Y, Schulten K.
\newblock Principal Component Analysis and Long Time Protein Dynamics.
\newblock J Phys Chem. 1996;100(7):2567--2572.

\bibitem{Kitao1999-jr}
Kitao A, Go N.
\newblock Investigating protein dynamics in collective coordinate space.
\newblock Curr Opin Struct Biol. 1999 Apr;9(2):164--169.

\bibitem{Huttenlocher1993-rr}
Huttenlocher DP, Klanderman GA, Rucklidge WJ.
\newblock Comparing images using the Hausdorff distance.
\newblock IEEE Trans Pattern Anal Mach Intell. 1993 Sep;15(9):850--863.

\bibitem{Alt1995-dh}
Alt H, Behrends B, Bl{\"{o}}mer J.
\newblock Approximate matching of polygonal shapes.
\newblock Ann Math Artif Intell. 1995 1~Sep;13(3-4):251--265.

\bibitem{Alt2008-lg}
Alt H, Scharf L.
\newblock Computing the {Hausdorff} distance between curved objects.
\newblock Int J Comput Geom Appl. 2008 Aug;18(04):307--320.

\bibitem{Frechet1906-ih}
Fr\'{e}chet M.
\newblock Sur quelques points du calcul fonctionnel.
\newblock Rend Circ Mat Palermo. 1906 Dec;22(1):1--72.

\bibitem{Alt1995-mc}
Alt H, Godau M.
\newblock Computing the {Fr\'{e}chet} distance between two polygonal curves.
\newblock Int J Comput Geom Appl. 1995 Mar;05(01n02):75--91.

\bibitem{Driemel2012-ji}
Driemel A, Har-Peled S, Wenk C.
\newblock Approximating the Fr\'{e}chet Distance for Realistic Curves in Near
  Linear Time.
\newblock Discrete Comput Geom. 2012 18~Jul;48(1):94--127.

\bibitem{Har-Peled2014-nx}
Har-Peled S, Raichel B.
\newblock The fr\'{e}chet distance revisited and extended.
\newblock ACM Trans Algorithms. 2014 1~Jan;10(1):3.

\bibitem{Eiter1994-wz}
Eiter T, Mannila H.
\newblock Computing Discrete {Fr\'{e}chet} Distance.
\newblock Wien: Christian Doppler Laboratory for Expert Systems, Technische
  Universit{\"{a}}t Wien; 1994.

\bibitem{Alt2001-wh}
{Helmut Alt, Christian Knauer,}.
\newblock Bounding the Fr\'{e}chet distance by the Hausdorff distance.
\newblock In: In Proceedings of the Seventeenth European Workshop on
  Computational Geometry; 2001. p. 166--169.

\bibitem{Buchin2008-tv}
Buchin K, Buchin M, Wenk C.
\newblock Computing the Fr\'{e}chet distance between simple polygons.
\newblock Comput Geom. 2008 Oct;41(1--2):2--20.

\bibitem{Lindorff-Larsen2009-wm}
Lindorff-Larsen K, Ferkinghoff-Borg J.
\newblock Similarity measures for protein ensembles.
\newblock PLoS One. 2009 15~Jan;4(1):e4203.

\bibitem{Sriraghavendra2007-lv}
Sriraghavendra R, Karthik K, Bhattacharyya C.
\newblock Fr\'{e}chet Distance Based Approach for Searching Online Handwritten
  Documents.
\newblock In: Document Analysis and Recognition, 2007. {ICDAR} 2007. Ninth
  International Conference on. vol.~1; 2007. p. 461--465.

\bibitem{De_Berg2011-kw}
de~Berg M, Cook AF IV.
\newblock Go with the Flow: The {Direction-Based} Fr\'{e}chet Distance of
  Polygonal Curves.
\newblock In: Marchetti-Spaccamela A, Segal M, editors. Theory and Practice of
  Algorithms in (Computer) Systems. vol. 6595 of Lecture Notes in Computer
  Science. Springer Berlin Heidelberg; 2011. p. 81--91.

\bibitem{Zhu2007-fl}
Zhu B.
\newblock Protein local structure alignment under the discrete Fr\'{e}chet
  distance.
\newblock J Comput Biol. 2007 Dec;14(10):1343--1351.

\bibitem{Jiang2008-fv}
Jiang M, Xu Y, Zhu B.
\newblock Protein structure-structure alignment with discrete Fr\'{e}chet
  distance.
\newblock J Bioinform Comput Biol. 2008 Feb;6(1):51--64.

\bibitem{Panchenko2004-gz}
Panchenko AR, Madej T.
\newblock Analysis of protein homology by assessing the (dis)similarity in
  protein loop regions.
\newblock Proteins. 2004 15~Nov;57(3):539--547.

\bibitem{Panchenko2005-qa}
Panchenko AR, Madej T.
\newblock Structural similarity of loops in protein families: toward the
  understanding of protein evolution.
\newblock BMC Evol Biol. 2005 3~Feb;5:10.

\bibitem{Jiang2014-zl}
Jiang W, Phillips JC, Huang L, Fajer M, Meng Y, Gumbart JC, et~al.
\newblock Generalized Scalable Multiple Copy Algorithms for Molecular Dynamics
  Simulations in {NAMD}.
\newblock Comput Phys Commun. 2014 Mar;185(3):908--916.

\bibitem{Dickson2012-dm}
Dickson BM, Huang H, Post CB.
\newblock Unrestrained computation of free energy along a path.
\newblock J Phys Chem B. 2012 13~Sep;116(36):11046--11055.

\bibitem{Gin2009-ho}
Gin BC, Garrahan JP, Geissler PL.
\newblock The limited role of nonnative contacts in the folding pathways of a
  lattice protein.
\newblock J Mol Biol. 2009 9~Oct;392(5):1303--1314.

\bibitem{Lenz2009-za}
Lenz P, Cho SS, Wolynes PG.
\newblock Analysis of single molecule folding studies with replica correlation
  functions.
\newblock Chem Phys Lett. 2009 26~Mar;471(4-6):310--314.

\bibitem{Graham2011-xp}
Graham TGW, Best RB.
\newblock Force-induced change in protein unfolding mechanism: discrete or
  continuous switch?
\newblock J Phys Chem B. 2011 17~Feb;115(6):1546--1561.

\bibitem{Lindorff-Larsen2011-wr}
Lindorff-Larsen K, Piana S, Dror RO, Shaw DE.
\newblock How fast-folding proteins fold.
\newblock Science. 2011 28~Oct;334(6055):517--520.

\bibitem{Huang2009-ws}
Huang H, Ozkirimli E, Post CB.
\newblock A Comparison of Three Perturbation Molecular Dynamics Methods for
  Modeling Conformational Transitions.
\newblock J Chem Theory Comput. 2009 9~Apr;5(5):1301--1314.

\bibitem{Ferrara2000-py}
Ferrara P, Apostolakis J, Caflisch A.
\newblock Computer simulations of protein folding by targeted molecular
  dynamics.
\newblock Proteins. 2000 15~May;39(3):252--260.

\bibitem{Ovchinnikov2012-pf}
Ovchinnikov V, Karplus M.
\newblock Analysis and elimination of a bias in targeted molecular dynamics
  simulations of conformational transitions: application to calmodulin.
\newblock J Phys Chem B. 2012 26~Jul;116(29):8584--8603.

\bibitem{Maheshwari2011-go}
Maheshwari A, Sack JR, Shahbaz K, Zarrabi-Zadeh H.
\newblock Fr\'{e}chet distance with speed limits.
\newblock Comput Geom. 2011 Feb;44(2):110--120.

\bibitem{Driemel2013-zm}
Driemel A, Har-Peled S.
\newblock Jaywalking Your Dog: Computing the {Fr\'{e}chet} Distance with
  Shortcuts.
\newblock SIAM J Comput. 2013;42(5):1830--1866.

\bibitem{Barton2010-dj}
Barto\v{n} M, Hanniel I, Elber G, Kim MS.
\newblock Precise Hausdorff distance computation between polygonal meshes.
\newblock Comput Aided Geom Des. 2010 Nov;27(8):580--591.

\bibitem{Krebs2000-di}
Krebs WG, Gerstein M.
\newblock The morph server: a standardized system for analyzing and visualizing
  macromolecular motions in a database framework.
\newblock Nucleic Acids Res. 2000 15~Apr;28(8):1665--1675.

\bibitem{Schulz1990-hp}
Schulz GE, M{\"{u}}ller CW, Diederichs K.
\newblock Induced-fit movements in adenylate kinases.
\newblock J Mol Biol. 1990 20~Jun;213(4):627--630.

\bibitem{Gerstein1993-jg}
Gerstein M, Schulz G, Chothia C.
\newblock Domain closure in adenylate kinase. Joints on either side of two
  helices close like neighboring fingers.
\newblock J Mol Biol. 1993 20~Jan;229(2):494--501.

\bibitem{Vonrhein1995-yq}
Vonrhein C, Schlauderer GJ, Schulz GE.
\newblock Movie of the structural changes during a catalytic cycle of
  nucleoside monophosphate kinases.
\newblock Structure. 1995 15~May;3(5):483--490.

\bibitem{Sinev1996-xp}
Sinev MA, Sineva EV, Ittah V, Haas E.
\newblock Domain closure in adenylate kinase.
\newblock Biochemistry. 1996 21~May;35(20):6425--6437.

\bibitem{Muller1996-gi}
M{\"{u}}ller CW, Schlauderer GJ, Reinstein J, Schulz GE.
\newblock Adenylate kinase motions during catalysis: an energetic counterweight
  balancing substrate binding.
\newblock Structure. 1996 15~Feb;4(2):147--156.

\bibitem{Henzler-Wildman2007-fb}
Henzler-Wildman KA, Lei M, Thai V, Kerns SJ, Karplus M, Kern D.
\newblock A hierarchy of timescales in protein dynamics is linked to enzyme
  catalysis.
\newblock Nature. 2007 6~Dec;450(7171):913--916.

\bibitem{Shapiro2006-xn}
Shapiro YE, Meirovitch E.
\newblock Activation energy of catalysis-related domain motion in E. coli
  adenylate kinase.
\newblock J Phys Chem B. 2006 15~Jun;110(23):11519--11524.

\bibitem{Hanson2007-wf}
Hanson JA, Duderstadt K, Watkins LP, Bhattacharyya S, Brokaw J, Chu JW, et~al.
\newblock Illuminating the mechanistic roles of enzyme conformational dynamics.
\newblock Proc Natl Acad Sci U S A. 2007 13~Nov;104(46):18055--18060.

\bibitem{Aden2007-kk}
Ad\'{e}n J, Wolf-Watz M.
\newblock {NMR} identification of transient complexes critical to adenylate
  kinase catalysis.
\newblock J Am Chem Soc. 2007 14~Nov;129(45):14003--14012.

\bibitem{Berman2000-eo}
Berman HM, Westbrook J, Feng Z, Gilliland G, Bhat TN, Weissig H, et~al.
\newblock The Protein Data Bank.
\newblock Nucleic Acids Res. 2000 1~Jan;28(1):235--242.

\bibitem{Muller1992-bq}
M{\"{u}}ller CW, Schulz GE.
\newblock Structure of the complex between adenylate kinase from Escherichia
  coli and the inhibitor {Ap5A} refined at 1.9 A resolution. A model for a
  catalytic transition state.
\newblock J Mol Biol. 1992 5~Mar;224(1):159--177.

\bibitem{Bennett1994-hn}
Bennett MJ, Choe S, Eisenberg D.
\newblock Domain swapping: entangling alliances between proteins.
\newblock Proc Natl Acad Sci U S A. 1994 12~Apr;91(8):3127--3131.

\bibitem{Bennett1994-um}
Bennett MJ, Choe S, Eisenberg D.
\newblock Refined structure of dimeric diphtheria toxin at 2.0 A resolution.
\newblock Protein Sci. 1994 Sep;3(9):1444--1463.

\bibitem{Bennett1994-im}
Bennett MJ, Eisenberg D.
\newblock Refined structure of monomeric diphtheria toxin at 2.3 resolution.
\newblock Protein Sci. 1994;3:1464--1475.

\bibitem{Humphrey1996aa}
Humphrey W, Dalke A, Schulten K.
\newblock {VMD} -- {V}isual {M}olecular {D}ynamics.
\newblock J Mol Graphics. 1996;14:33--38.
\newblock Available from: \url{http://www.ks.uiuc.edu/Research/vmd/}.

\bibitem{Dahl:2012ap}
Dahl ACE, Chavent M, Sansom MSP.
\newblock Bendix: Intuitive helix geometry analysis and abstraction.
\newblock Bioinformatics. 2012;28(16):2193--2194.

\bibitem{Hunter:2007aa}
Hunter JD.
\newblock Matplotlib: A 2D graphics environment.
\newblock Computing In Science \& Engineering. 2007 May-Jun;9(3):90--95.

\bibitem{Waskom2014:aa}
Waskom M, Botvinnik O, Hobson P, Cole JB, Halchenko Y, Hoyer S, et~al..
  seaborn: v0.5.0 (November 2014); 2014.
\newblock Available from: \url{http://dx.doi.org/10.5281/zenodo.12710}.

\bibitem{Hintze:1998tw}
Hintze JL, Nelson RD.
\newblock Violin Plots: A Box Plot-Density Trace Synergism.
\newblock The American Statistician. 1998;52(2):181--184.

\bibitem{Michaud-Agrawal2011-yg}
Michaud-Agrawal N, Denning EJ, Woolf TB, Beckstein O.
\newblock {MDAnalysis}: a toolkit for the analysis of molecular dynamics
  simulations.
\newblock J Comput Chem. 2011 30~Jul;32(10):2319--2327.

\bibitem{Xu2008-mg}
Xu R, Wunsch D.
\newblock Clustering.
\newblock IEEE Press Series on Computational Intelligence. John Wiley \& Sons;
  2008.

\bibitem{Ward1963-mp}
Ward JH Jr.
\newblock Hierarchical Grouping to Optimize an Objective Function.
\newblock J Am Stat Assoc. 1963 Mar;58(301):236--244.

\bibitem{SciPy}
Jones E, Oliphant T, Peterson P, et~al.. {SciPy}: Open source scientific tools
  for {Python}; 2001--.
\newblock [Online; accessed 2015-05-13].
\newblock Available from: \url{http://www.scipy.org/}.

\bibitem{Shakhnovich1991-ae}
Shakhnovich E, Farztdinov G, Gutin AM, Karplus M.
\newblock Protein folding bottlenecks: A lattice Monte Carlo simulation.
\newblock Phys Rev Lett. 1991 16~Sep;67(12):1665--1668.

\bibitem{Teodoro2003-lq}
Teodoro ML, Phillips GN Jr, Kavraki LE.
\newblock Understanding protein flexibility through dimensionality reduction.
\newblock J Comput Biol. 2003;10(3-4):617--634.

\bibitem{Mesentean2006-gl}
Mesentean S, Fischer S, Smith JC.
\newblock Analyzing large-scale structural change in proteins: comparison of
  principal component projection and Sammon mapping.
\newblock Proteins. 2006 1~Jul;64(1):210--218.

\bibitem{Beckstein2009-ll}
Beckstein O, Denning EJ, Perilla JR, Woolf TB.
\newblock Zipping and Unzipping of Adenylate Kinase: Atomistic Insights into
  the Ensemble of {Open $\leftrightarrow$ Closed} Transitions.
\newblock J Mol Biol. 2009 Nov;394(1):160--176.

\bibitem{Sfriso2012-uf}
Sfriso P, Emperador A, Orellana L, Hospital A, Gelp\'{\i} JL, Orozco M.
\newblock Finding Conformational Transition Pathways from Discrete Molecular
  Dynamics Simulations.
\newblock J Chem Theory Comput. 2012;8(11):4707--4718.

\bibitem{Sfriso2013-ah}
Sfriso P, Hospital A, Emperador A, Orozco M.
\newblock Exploration of conformational transition pathways from coarse-grained
  simulations.
\newblock Bioinformatics. 2013 15~Aug;29(16):1980--1986.

\bibitem{Das2014-ry}
Das A, Gur M, Cheng MH, Jo S, Bahar I, Roux B.
\newblock Exploring the conformational transitions of biomolecular systems
  using a simple two-state anisotropic network model.
\newblock PLoS Comput Biol. 2014 Apr;10(4):e1003521.

\bibitem{Tekpinar2010-kc}
Tekpinar M, Zheng W.
\newblock Predicting order of conformational changes during protein
  conformational transitions using an interpolated elastic network model.
\newblock Proteins. 2010 15~Aug;78(11):2469--2481.

\bibitem{Zheng2007-es}
Zheng W, Brooks BR, Hummer G.
\newblock Protein conformational transitions explored by mixed elastic network
  models.
\newblock Proteins. 2007 1~Oct;69(1):43--57.

\bibitem{Perilla2011-im}
Perilla JR, Beckstein O, Denning EJ, Woolf TB.
\newblock Computing ensembles of transitions from stable states: Dynamic
  importance sampling.
\newblock J Comput Chem. 2011 30~Jan;32(2):196--209.

\bibitem{Vanden-Eijnden:2009dq}
Vanden-Eijnden E, Venturoli M.
\newblock Revisiting the finite temperature string method for the calculation
  of reaction tubes and free energies.
\newblock J Chem Phys. 2009 May;130(19):194103.

\bibitem{Huber:1996ht}
Huber G, Kim S.
\newblock Weighted-ensemble Brownian dynamics simulations for protein
  association reactions.
\newblock Biophys J. 1996 Jan;70(1):97--110.

\bibitem{Zhang:2010ve}
Zhang BW, Jasnow D, Zuckerman DM.
\newblock The "weighted ensemble" path sampling method is statistically exact
  for a broad class of stochastic processes and binning procedures.
\newblock J Chem Phys. 2010 Feb;132(5):054107.

\bibitem{Bello-Rivas:2015vn}
Bello-Rivas JM, Elber R.
\newblock Exact milestoning.
\newblock J Chem Phys. 2015 Mar;142(9):094102.

\bibitem{Bolhuis:2002qe}
Bolhuis P, Chandler D, Dellago C, Geissler P.
\newblock Transition path sampling: {Throwing} ropes over rough mountain
  passes, in the dark.
\newblock Ann Rev Phys Chem. 2002;53:291--318.

\bibitem{Warmflash:2007kl}
Warmflash A, Bhimalapuram P, Dinner AR.
\newblock Umbrella sampling for nonequilibrium processes.
\newblock J Chem Phys. 2007 Oct;127(15):154112.

\bibitem{Allen:2006cr}
Allen RJ, Frenkel D, ten Wolde PR.
\newblock Simulating rare events in equilibrium or nonequilibrium stochastic
  systems.
\newblock J Chem Phys. 2006 Jan;124(2):024102.

\bibitem{Brooks2009-be}
Brooks BR, Brooks~III CL, Mackerell ADJ, Nilsson L, Petrella RJ, Roux B, et~al.
\newblock {{CHARMM}}: the biomolecular simulation program.
\newblock J Comput Chem. 2009 Jul;30(10):1545--1614.

\bibitem{Phillips2005-hx}
Phillips JC, Braun R, Wang W, Gumbart J, Tajkhorshid E, Villa E, et~al.
\newblock Scalable molecular dynamics with {NAMD}.
\newblock J Comput Chem. 2005 Dec;26(16):1781--1802.

\bibitem{MacKerell98}
MacKerell A, Bashford D, Bellott M, Dunbrack R, Evanseck J, Field M, et~al.
\newblock All-atom empirical potential for molecular modeling and dynamics
  studies of proteins.
\newblock J Phys Chem B. 1998;102(18):3586--3616.

\bibitem{MacKerell04a}
MacKerell AD Jr, Feig M, {Brooks III} CL.
\newblock Extending the treatment of backbone energetics in protein force
  fields: limitations of gas-phase quantum mechanics in reproducing protein
  conformational distributions in molecular dynamics simulations.
\newblock J Comput Chem. 2004;25:1400--1415.

\bibitem{Schaefer:2001et}
Schaefer M, Bartels C, Leclerc F, Karplus M.
\newblock Effective atom volumes for implicit solvent models: comparison
  between {Voronoi} volumes and minimum fluctuation volumes.
\newblock J Comput Chem. 2001 Nov;22(15):1857--1879.

\bibitem{Tanner:2011oq}
Tanner DE, Chan KY, Phillips JC, Schulten K.
\newblock Parallel Generalized Born Implicit Solvent Calculations with NAMD.
\newblock Journal of Chemical Theory and Computation. 2011;7(11):3635--3642.

\bibitem{Flores2006-bf}
Flores S, Echols N, Milburn D, Hespenheide B, Keating K, Lu J, et~al.
\newblock The Database of Macromolecular Motions: new features added at the
  decade mark.
\newblock Nucleic Acids Res. 2006 1~Jan;34(Database issue):D296--301.

\bibitem{Maiorov1994-od}
Maiorov VN, Crippen GM.
\newblock Significance of root-mean-square deviation in comparing
  three-dimensional structures of globular proteins.
\newblock J Mol Biol. 1994 14~Jan;235(2):625--634.

\bibitem{Buchin:2011zr}
Buchin K, Buchin M, van Kreveld M, Luo J.
\newblock Finding long and similar parts of trajectories.
\newblock Computational Geometry. 2011;44(9):465 -- 476.

\bibitem{Sankararaman:2013ys}
Sankararaman S, Agarwal PK, M{\o}lhave T, Pan J, Boedihardjo AP.
\newblock Model-driven matching and segmentation of trajectories.
\newblock In: 21st {SIGSPATIAL} International Conference on Advances in
  Geographic Information Systems, {SIGSPATIAL} 2013, Orlando, FL, USA, November
  5-8, 2013; 2013. p. 234--243.

\end{thebibliography}

\end{document}